\documentclass[aps,prc,showpacs,nofootinbib,preprint]{revtex4}
\usepackage{amsmath}
\usepackage{amssymb}
\usepackage{graphics}
\begin{document}
\title{Unexpected impact of $D$ waves in low-energy neutral pion photoproduction from the proton
and the extraction of multipoles}

\author{C.~\surname{Fern\'andez-Ram\'{\i}rez}}
\email{ramirez@ect.it}
\altaffiliation{Current address: European Centre for Theoretical Studies in Nuclear Physics and Related Areas (ECT*), Strada delle Tabarelle 286, I-38050 Villazzano (TN), Italy.}
\affiliation{Center for Theoretical Physics, Laboratory for Nuclear Science and Department of Physics,
Massachusetts Institute of Technology,
77 Massachusetts Ave., Cambridge, Massachusetts  02139, USA}
\author{A.M.~\surname{Bernstein}} 
\affiliation{Center for Theoretical Physics, Laboratory for Nuclear Science and Department of Physics,
Massachusetts Institute of Technology,
77 Massachusetts Ave., Cambridge, Massachusetts  02139, USA}
\author{T.W.~\surname{Donnelly}}
\affiliation{Center for Theoretical Physics, Laboratory for Nuclear Science and Department of Physics,
Massachusetts Institute of Technology,
77 Massachusetts Ave., Cambridge, Massachusetts  02139, USA}

\date{\today}

\begin{abstract}
Contributions of $D$ waves to physical observables for neutral
pion photoproduction from the proton in the near-threshold region are studied
and means to isolate them are proposed.
Various approaches to describe the multipoles are employed
 ---a phenomenological one, a unitary one, and heavy baryon chiral perturbation theory. 
The results of these approaches are compared and found to yield essentially the same
answers.
$D$ waves are seen to enter together with $S$ waves in a way that any means which 
attempt to obtain the
$E_{0+}$ multipole accurately must rely on knowledge of $D$ waves
and that consequently the latter
cannot be dismissed in analyses of low-energy pion photoproduction. 
It is shown that $D$ waves have a significant impact on 
double-polarization observables that can be measured.
This importance of $D$ waves is due to the soft nature of the $S$ wave
and is a direct consequence of chiral symmetry and the Nambu--Goldstone nature of the pion.
$F$-wave contributions are shown to be negligible
in the near-threshold region.

\end{abstract}
\pacs{12.39.Fe, 13.60.Le, 25.20.Lj}
\maketitle

\section{Introduction}
Due to the spontaneous breaking of chiral symmetry in Quantum Chromodynamics (QCD) the $\pi$ meson appears as a pseudoscalar Nambu--Goldstone boson \cite{book}. 
As a dynamical consequence,
the $S$-wave amplitude for the $\gamma N \rightarrow \pi^{0} N$ reaction is small in the 
threshold region, because it vanishes in the chiral limit, \textit{i.e.} when the light quark masses 
are set equal to zero 
\cite{CHPT91,CHPT96,CHPT01}. An additional  consequence is that the 
$P$-wave amplitude is large and leads to the $\Delta$ resonance at intermediate
energies \cite{AB-Delta}.
Accordingly,  the photoproduction of neutral pions differs from the general pattern 
for hadronic reactions where
the $S$ wave dominates close to threshold and then, as the energy increases, the higher angular momentum waves ($P$, $D$, \ldots)  start to become important. By contrast, for the 
$\gamma N \rightarrow \pi^{0} N$ reaction the $S$- and $P$-wave contributions
are comparable  even very close to threshold \cite{AB-fits}.
Hence, the accurate extraction of  $S$ and $P$ waves from pion photoproduction data becomes
an important issue in the study of the breaking of chiral symmetry and QCD.

The partial waves (electromagnetic multipoles) 
are not experimental observables, but rather are quantities extracted from data
using some kind of approach and/or theoretical input 
--- Heavy Baryon Chiral Perturbation Theory (HBCHPT), for instance.

In the first resonance region ($\Delta$ region) the influence of higher partial waves was addressed 
in \cite{dwavesindelta}  employing the energy-dependent solution
of SAID \cite{SAID}; however when it comes to precision physics in the near-threshold region,
the standard approach consists in analyzing data assuming that the partial waves 
of interest ($S$ and $P$ waves) are sufficient to
describe the experimental data so, higher partial waves may be neglected
\cite{CHPT91,CHPT96,CHPT01,AB-fits,Schmidt,AB-review}.
In the literature there are several phenomenological 
models of neutral pion photoproduction based on
the standard Born terms \cite{Peccei} that
are applied at energies spanning from threshold to the first resonance region.
These models compute the electromagnetic multipoles and include higher partial waves in their analysis, although they are not adequate
to address precision physics in the near-threshold region due to model dependencies 
(e.g. form factors and resonances treatment) and large uncertainties in the multipoles.
As an example, the most accurate model calculation in the near-threshold energy region is the 
Dubna--Mainz--Taipei (DMT) dynamical model \cite{DMT}
whose $P$ waves do not provide a good description of the experimental data for the
polarization asymmetry.
Two arguments have motivated this assumption in previous analyses: 
(i) The angular dependence of the experimental differential cross section
can be described accurately using  Legendre polynomials up only to order two.
Indeed, $S$ and $P$ waves constitute the minimal set of partial waves needed to reach that angular dependence;
(ii) In the near-threshold energy region higher partial waves are weak
and the early dominance of the $M_{1+}$ multipole renders them negligible.

As will be discussed in Sec. \ref{sec:generalstructure},
the first statement is misleading because any coefficient that accompanies a Legendre polynomial in the expansion depends on \textit{all} partial waves,
implying that in some circumstances 
higher partial waves can make an important contribution to the coefficients
that accompany the lower-order Legendre polynomials, 
thereby posing uncertainties in the multipole extraction.
This assumption cannot be taken for granted and has to be tested.
The second argument is actually in favor  of 
the possible importance of higher partial waves and not against it, as is usually stated. 
The conclusion disregards that the $S$ wave is also weak
and that important features of the angular dependence of the observables
are dominated by the interference of different partial waves. In this situation the dominance of a certain contribution (such the $M_{1+}$ in this case) can lead to an important enhancement of smaller partial waves through interference, making them relevant.
An excellent example of this situation is the well-known $S$-wave/$D$-wave interference in the beam asymmetry in $\eta$ photoproduction, where a very weak $D$ wave turns out to be necessary to explain the experimental data due to its interference with the dominant $S$ wave.
In this article it is shown that something similar, although more subtle, happens in the case of pion photoproduction, where $D$ waves affect the extraction of the $S$ wave.

The purpose of this article is twofold: 1) to complement and extend the analysis of what was reported in Ref. \cite{FBD09} where the impact of $D$-wave contributions in the near-threshold region was explored
using as starting point HBCHPT  and
2) to study the impact of the $D$-wave contribution on the observables and assess the measurability of
this impact.

As  will be seen in next sections  
the understanding of $D$ waves and their impact on the observables is necessary if 
one hopes to extract the $S$ wave accurately and to arrive at conclusions regarding isospin breaking, unitarity, energy dependence of the multipole, the magnitude of the unitary cusp, or the extraction of the Low Energy Constants (LECs) in HBCHPT.

\section{General structure of the observables in terms of the electromagnetic multipoles} \label{sec:generalstructure}

It has been customary to study pion photoproduction from the nucleon in the near threshold energy region (up to approximately 170 MeV) using only the contributions of $S$ and $P$ waves and there is abundant literature that discusses the structure of the observables in terms of only these partial waves \cite{CHPT91,CHPT96,CHPT01,AB-fits,AB-review,DonnellyRaskin} ---namely, 
using $E_{0+}$, $E_{1+}$, $M_{1-}$, and $M_{1+}$ electromagnetic multipoles. In Ref. \cite{FBD09} it was found that $D$ waves play an unexpected role in this energy region and that $D$ waves are required to  extract the $E_{0+}$ multipole accurately.

\begin{table*}
\caption{Relationships among several notations present in the literature for the asymmetries and responses. In this article we use the sign conventions of Ref. \cite{CT97} for the asymmetries. The first line contains different notations for the differential cross section and the associated response.
The second line stands for the photon beam asymmetry, measured as the difference between the differential cross section for perpendicular and parallel polarized photons.
The third line contains several conventions for the beam-target asymmetry obtained by measuring 
the difference of the cross sections for the two circular polarizations of the photon with a target  polarized along the beam direction.
The fourth line contains several conventions for the beam-target asymmetry obtained by measuring 
the difference of the cross sections for the two circular polarizations of the photon, but now
with target polarized in the sideways direction.
In the second column, $\vec{\gamma}$ stands for linear polarization of the photon while $\gamma_c$
stands for circular polarization of the photon. The $z$ stands for longitudinal polarization of the target (along the photon direction) and the $x$ for sideways polarization of the target (transverse and in the scattering plane).
In the third column, $0$ in $W^T(0)$ and $W^{TT}(0)$ stands for unpolarized target.
The $s$ in $W^{T'}(s)$ stands for target polarized in the sideways direction (in any other place in this article $s$ stands for the Mandelstam variable) and the $l$ in $W^{T'}(l)$ for target polarized in the beam direction.} \label{tab:notation}
\begin{ruledtabular}
\begin{tabular}{cccc}
 This article  &  Ref. \cite{AB-review}  & Ref. \cite{DonnellyRaskin} &   Ref. \cite{BDS75}   \\
 \hline
$\sigma_T \equiv q_\pi W_T /k_\gamma$ ; Eq. (\ref{eq:eq1})    &$\sigma_T \equiv q_\pi R^{00}_{T}/k_\gamma$ & $q_\pi W^T(0)/ k_\gamma$ &    $d\sigma/d\Omega$\\
$\Sigma \equiv-W_S/W_T$ ;  Eq. (\ref{eq:eq2})    &$A(\vec{\gamma}) \equiv-R^{00}_{TT}/R^{00}_{T}$  & $-W^{TT}(0)/W^T(0)$ &   $\Sigma$ \\
$E \equiv W_E/W_T$ ;  Eq. (\ref{eq:eq3}) &$A(\gamma_c,z) \equiv -R^{0z}_{TT'}/R^{00}_{T}$  & $-W^{T'}(l)/W^T(0)$ &   $E$ \\
$F \equiv W_F/W_T$  ;  Eq. (\ref{eq:eq4}) &$-A(\gamma_c,x) \equiv -R^{0x}_{TT'}/R^{00}_{T}$  &$W^{T'}(s)/W^T(0)$ & $-F$
\end{tabular}
\end{ruledtabular}
\end{table*}

This article expands the work reported in Ref. \cite{FBD09}  and the first step consists in analyzing the structure of the observables in terms of the reported electromagnetic multipoles when the $D$ waves are added.
All photoproduction observables 
can be written in terms of responses of two kinds, time reversal even (TRE) and time reversal odd (TRE).
Both types of
responses are real quantities obtained by taking either the real or imaginary part of bilinear products of the complex multipoles, respectively. 
If the real part is chosen, the observable is TRE ($\sigma_T$, $\Sigma$, $E$, $F$, $C_x$, $C_z$, $T_x$, $T_z$, $L_x$, and $L_z$) and if the imaginary part is chosen the observable is TRO ($T$, $P$, $G$, $H$, $O_x$, $O_z$). See Refs. \cite{BDS75,CT97} for the definition of the observables.
From this full set of observables, in this article the focus is placed on those that do not require one to measure the recoil polarization, in other words, those that  involve polarizing only the beam, the target, both  or neither, namely
the ones that could be accessed experimentally in the near-threshold region.
That leaves  four TRE---$\sigma_T$, $\Sigma$, $E$, and $F$---and
four TRO---$T$, $P$, $G$, and $H$---observables, which will be discussed in more detail in the following.
In Sec. \ref{sec:results} a selection of several
observables at different energies and angles is presented, restricting the discussion (and the tables in Appendix  \ref{sec:apendiceA}) to the most promising  ones involved in the search for $D$ waves and their interplay with $S$ and $P$ waves,
namely, the differential cross section $\sigma_T$, the photon beam asymmetry $\Sigma$, and the double-polarization asymmetries $E$ and $F$ (circular photon polarization and  polarization of the target, respectively, along the beam axis and the orthogonal axis within the scattering plane).
For the nomenclature used see Table \ref{tab:notation}; 
the structure of these observables is provided in Appendix \ref{sec:apendiceA}.
All these observables have become accessible experimentally in recent years thanks to the existence of high-duty-cycle photon facilities such as MAMI (Mainz) and HI$\gamma$S (Duke) and to the development of polarized targets.

Any pion photoproduction response $\mathcal{R} \left(s,\theta \right)$ depending on
Mandelstam variable $s$ and photon-pion angle $\theta$ can be expanded in terms of Legendre polynomials  $P_j \left( \theta \right)$ times $\sin\theta$ to a specific integer power $n$:
\begin{equation}
\mathcal{R} \left(s,\theta \right) = \sin^n \theta \left[ \: \mathcal{R}_0\left(s \right) + \mathcal{R}_1 \left(s \right) \mathcal{P}_1 \left( \theta \right)  + \dots \right] \: ,
\end{equation}
where coefficients $\mathcal{R}_j\left(s \right)$ can be defined on terms of the electromagnetic multipoles up to a certain partial wave and the number of Legendre polynomials also depends on up to which partial wave the observable is expanded. One has $n=0$ for $W_T$ and $W_E$, $n=1$ for $W_F$ and $n=2$ for $W_S$.

For example, the differential cross section can be written including up to $D$ waves as:
\begin{equation}
\begin{split}
\sigma_T (s,\theta)= \frac{q_\pi}{k_\gamma} \left[
T_0 \left( s \right)+ T_1\left( s \right) \mathcal{P}_1\left( \theta \right)  
+ T_2 \left( s \right) \mathcal{P}_2\left( \theta \right) \right. \\
\left. + T_3\left( s \right) \mathcal{P}_3\left( \theta \right) 
+ T_4 \left( s \right) \mathcal{P}_4\left( \theta \right) \right] \: , 
\end{split} \label{eq:sigma}
\end{equation}
where $q_\pi$ and $k_\gamma$ are the pion and photon momenta in the center of mass, respectively,
and the $T_i$ depend on the photon energy.

The full calculation of the responses up to $D$ waves can be found in Ref. \cite{Knochlein95}, although 
here we have preferred to proceed using the developments in Ref.
\cite{DonnellyRaskin} and to present the results using tables 
instead of lengthly equations (see Appendix \ref{sec:apendiceA}).
The reason for using tables will become apparent later when it will be seen  that they make
it easier to extract the interplay of the interferences between partial waves.

The first place where effects of $D$ waves in low-energy pion photoproduction were found \cite{FBD09} was the part of the differential cross section that is associated with the Legendre polynomial $\mathcal{P}_{1}\left( \theta \right)$, namely $T_1$ in Eq. (\ref{eq:sigma}).
This is used here as a working example of how to employ the tables.
From Eq. (\ref{eq:Tn}) this piece of the response can be written:
\begin{equation}
T_1 \left( s \right)=\sum_{ij} \text{Re} \{ \: \mathcal{M}^*_i \left( s \right) \: T_1^{ij} \: \mathcal{M}_j \left( s \right) \: \} \: ,
\end{equation}
where
$\mathcal{M}_j \left( s \right) =E_{0+}$, $E_{1+}$, $E_{2+}$, $E_{2-}$, $M_{1+}$, $M_{1-}$, $M_{2+}$, $M_{2-}$, and the coefficients $T_1^{ij}$ can be read from Table \ref{tab:T1} 
in Appendix \ref{sec:apendiceA}, obtaining
\begin{equation}
T_1= 2 \: \text{Re} \left[ P_1^* E_{0+} \right] + \delta T_1 \: ,
\end{equation}
where $P_1\equiv 3E_{1+}+M_{1+}-M_{1-}$, 
and $\delta T_1$ stands for the $D$-wave/$P$-wave interference contribution
\begin{equation}
\begin{split}
\delta T_1 &= 2 \: \text{Re} \Big{[} \: \frac{27}{5}M^*_{1+}M_{2+} 
+\left( M^*_{1+} - M^*_{1-} \right) E_{2-} \\
&+E^*_{1+} \left( \frac{72}{5}E_{2+}-\frac{3}{5}E_{2-} +\frac{9}{5}M_{2+}
-\frac{9}{5}M_{2-} \right) \\
&+ \left( \frac{3}{5}M^*_{1+}+3M^*_{1-} \right) M_{2-}   \Big{]} \: .
\end{split}
\end{equation}

Proceeding in a similar way, the other observables and multipolar expansions can be worked out.
For instance, one can analyze the structure of the differential cross section in order to discover in which observables $D$ waves may show up.
If one takes into account only $S$ and $P$ waves, then only $T_0$, $T_1$, and $T_2$ coefficients contribute
to Eq. (\ref{eq:sigma}). 
If we add $D$ waves, two more quantities appear, $T_3$ and $T_4$.
So the first place to look for $D$ waves consists of checking to see if there is room for the appearance
of these new terms. The currently available experimental data \cite{Schmidt}
can be described quite well using only $T_0$, $T_1$, and $T_2$, 
and no $T_3$ or $T_4$ contribution appears to be required at present in the near-threshold region.
Hence, any contribution of $D$ waves to $\sigma_T  (s,\theta)$ should appear only as a modification of
$T_0$, $T_1$ or $T_2$. On the other hand, $T_0$ and $T_2$ are dominated by diagonal terms
involving the multipoles (which can be immediately read from Tables \ref{tab:T0} and \ref{tab:T2}), namely 
$|M_{1+}|^2$, $|M_{1-}|^2$, and $|E_{0+}|^2$, and thus any interference with $D$ waves would be negligible compared with the leading-order terms.
On the other hand, $T_1$ is entirely due to multipole interferences (all the coefficients in Table \ref{tab:T1} are off-diagonal), and so any $D$-wave interference with the dominant $M_{1+}$ multipole is a candidate for a non-negligible $D$-wave  contribution to the observable.
This affects any multipolar extraction from data using only $S$ and $P$ partial waves as was shown
in Ref. \cite{FBD09}.

It is very important to realize that without a so-called complete experiment ---which at this point is not feasible in the near-threshold region ---the multipole extraction from experimental data 
depends on the approach employed. 
For this reason we explore three prescriptions for the multipoles,
with different phenomenological and theoretical content. 
The first one is a pure phenomenological approach where the energy dependence is prescribed for the multipoles (embedded in Sec. \ref{sec:general}). 
The second consists in applying HBCHPT, which is the theoretically soundest approach
to computing the $S$- and $P$-wave multipoles (Sec. \ref{sec:hbchpt}). The last one uses HBCHPT to compute the $P$ waves, but a unitary prescription is employed for $E_{0+}$ (Sec. \ref{sec:unitary})
---the reasons for this choice will be detailed in the corresponding section.

For all three approaches the $D$ waves are computed using 
the customary Born terms (equivalent to the Born contribution to HBCHPT)
and vector-meson exchange ($\omega$ and $\rho$) 
\cite{FMU06}.
For the $\omega$ and $\rho$ parameters we have used
those given by the dispersion analysis of the form factors in Ref. \cite{MMD} 
that agrees with the latest analysis in Ref. \cite{Belushkin}.
The vector-meson correction is very small and the inclusion of $D$ waves in this way is almost 
equivalent to the zeroth order in HBCHPT.
For all approaches we perform fits using either solely $S$ and $P$ waves 
(SP fits) or  $S$, $P$ and $D$ waves (SPD fits).
The inclusion of the vector mesons in this fashion poses a model uncertainty in the calculation;
however, this uncertainty is very small and the conclusions on the effect of $D$ waves in the observables and the extraction of the multipoles is effectively model independent.

Within this framework we have performed fits
to the latest experimental data from MAMI \cite{Schmidt} 
(171 differential cross sections and 7 photon asymmetries, spanning the energy range from threshold up to 166 MeV).
As a fitting procedure
we have used a hybrid optimization code based on a genetic algorithm (GA)
combined with the 
E04FCF routine from the NAG library \cite{NAG}. 
In recent years, increased
credit is being given in nuclear and particle physics
to modern optimization procedures
\cite{OPTIMIZACION,PRC08}
and the error analysis for the parameters resulting from the fits.
Modern and sophisticated optimization techniques 
such as GAs \cite{Goldberg}
have been developed
over the past twenty years and have been applied to problems that were
impossible to face with conventional tools.
Although GAs are computationally more expensive 
than other algorithms, in a minimization problem they are much 
less likely to get stuck in local minima than are other approaches such as
gradient-based minimization methods, and they
allow one to explore a large parameter space more efficiently.
Thus, in a multiparameter minimization as the one we face here,
GAs probably provide the best possibility in searching for the minimum.
Moreover, they provide
additional information on the local minima structure.
The details of the fitting procedure, its technicalities
and advantages can be found in Ref. \cite{PRC08}.

\section{The electromagnetic multipoles}

\subsection{Unitarity and the General Form of the Multipoles} \label{sec:general}

From general principles such as time reversal invariance and unitarity
the $S$ wave can be written as the combination of a smooth part and a cusp part \cite{AB-lq,Anant}
\begin{equation}
\begin{split}
E_{0+} =& e^{i\delta_{0}} \left[ A_{0} + i \beta q_{+} /m_{\pi^+} \right]  \, ; \, W>W_{thr}(\pi^{+}n) \\
E_{0+} =& e^{i\delta_{0}} \left[ A_{0}  -  \beta  \left| q_{+} \right| /m_{\pi^+} \right] \, ; \, W<W_{thr}(\pi^{+}n) \: ,
\end{split} \label{eq:FW}
\end{equation}
where $\delta_0$ is the $\pi^0p$ phase shift (which is very small), $W$ is the invariant mass,
$W_{thr}(\pi^{+}n)$ the invariant mass at the $\pi^{+}n$ threshold,
$q_{+}$ is the $\pi^{+}$ center-of-mass momentum, $A_{0}$ is $E_{0+}$ in the absence of the 
charge exchange re-scattering (smooth part), and $\beta$ parameterizes the magnitude of the unitary cusp and  can be calculated \cite{AB-lq} on the basis of unitarity.
Equations (\ref{eq:FW}) provide a generalization of the Fermi--Watson theorem \cite{FW} by removing the requirement of isospin conservation; they have been derived using a three-coupled-channel 
$S$-matrix approach in which unitarity and time reversal invariance are satisfied \cite{AB-lq,Anant}. 
These calculations take the static isospin breaking (mass differences)  as well as $\pi$N scattering to all orders into account. In the electromagnetic sector they have been carried out to first order in the fine structure constant $\alpha$. The resulting equation for $\beta$ is 
\begin{equation}
\beta = E_{0+}(\gamma p \to \pi^{+} n)  \times  a(\pi^{+} n \to \pi^{0} p) \: , \label{eq:beta}
\end{equation}
where $a(\pi^{+} n \to \pi^{0} p)$ is the $\pi N$ $S$-wave charge exchange scattering length.
The value of $\beta$ will be discussed in Sec. \ref{sec:resultsA}.

The smooth part of $E_{0+}$ can be approximated by using a Taylor expansion, and hence the 
entire $S$-wave multipole can be written:
\begin{equation}
E_{0+} = E_{0+}^{(0)} + E_{0+}^{(1)} \frac{k^L_\gamma-k_\gamma^{T}}{m_{\pi^0}} 
+ i\beta \frac{q_{\pi^+}}{m_{\pi^+} } \: ,  \label{eq:Swave}
\end{equation}
where $E_{0+}^{(0)}$ and $E_{0+}^{(1)}$ are free parameters, $k^L_\gamma$ is the photon energy in laboratory frame, and $k_\gamma^{T}=144.681$ MeV  is the photon energy at threshold in the laboratory frame. The $\pi^+$ center-of-mass momentum, $q_{\pi^+}$, is real above and
imaginary below the $\pi^+$ threshold; this is a unitary cusp.
HBCHPT matches this expansion for the $S$ wave. 

Another expression for the $E_{0+}$ multipole is \cite{CHPT96,CHPT01,VKM2005}
\begin{equation}
E_{0+}=a+b\sqrt{1- E^2_\gamma/E^2_\text{c}} \: , \label{eq:Smeissner}
\end{equation}
where $a$ and $b$ are fitted to data, $E_\gamma$ is the center-of-mass photon energy and 
$E_\text{c}$ the center-of-mass photon energy at the $\pi^+$ production threshold.
Equation (\ref{eq:Smeissner}) is nearly equal to Eq. (\ref{eq:Swave}) at the 3.5\% level or better 
if $b=\beta$ and  $E^{(1)}_{0+}=0$. 
Due to this latter restriction we do not use Eq. (\ref{eq:Smeissner}) in this work.

The $P$ waves can also be studied expanding in a Taylor series and assuming the multipoles are real 
we only need the lowest two orders:
\begin{equation}
\frac{P_i}{q_{\pi}/m_{\pi^0}} = P_i^{(0)} + P_i^{(1)} \frac{k^L_\gamma-k_\gamma^{T}}{m_{\pi^0}} \: \: \text{;} \: \: i=1,2,3 \: , \label{eq:Pwaves}
\end{equation}
where $q_{\pi}$ is the outgoing pion momentum and
$P_i^{(0)}$ and $P_i^{(1)}$ are the coefficients of the expansion (free parameters). 
This expansion matches the energy dependence of HBCHPT for the real part of the multipoles.
The $P_i$ waves are related to the standard electromagnetic multipoles through
\begin{eqnarray}
E_{1+}  &=& \left( P_1 + P_2 \right)/6  \\
M_{1+} &=&  \left( P_1 - P_2 \right)/6 + P_3/3   \\
M_{1-}  &=& \left( P_3 + P_2 - P_1 \right)/3 
\end{eqnarray}

Another choice for the $P$ waves is the one made in Schmidt \textit{et al.} \cite{Schmidt}, where the prescription $P_i= P_i^{eff} k_\gamma q_{\pi}/m^2_{\pi^0}$ was employed where $P_i^{eff}$ is a constant fitted to data.
We do not use such a prescription because it does not match the energy dependence of HBCHPT near threshold \cite{CHPT96} and it is not truly the lowest-order Taylor expansion.
The $P$-wave expansion is divided by $q_{\pi}$ due to the angular momentum barrier.
The $D$-wave expansion, to first order, is
\begin{equation}
D_i = D_i^{(0)} q_{\pi}^2/m^2_{\pi^0} \: , \label{eq:Dwaves}
\end{equation}
where $D_i=E_{2+},E_{2-},M_{2+},M_{2-}$, and $q_{\pi}^2$ accounts for the correct angular 
momentum barrier. 
In this study the $D$ waves are computed using Born terms and vector mesons that
closely approximates this behaviour.

Even fixing the $D$ waves, these simple phenomenological prescriptions for the partial waves have too many parameters describing the energy dependence to perform a unique fit of the data.  The reason is that the $\Sigma$ (polarized photon asymmetry) measurement has only been published at one energy. However, a more extensive asymmetry data set has been obtained in experiments recently performed at MAMI and the data analysis is in progress \cite{Hornidge}. Therefore, at this time a phenomenological analysis is not possible and one has to rely on theoretical approaches such as the ones we employ in the rest of the article.

\subsection{Heavy Baryon Chiral Perturbation Theory} \label{sec:hbchpt}

At present, the best available theoretical framework for studying
pion photoproduction in the near-threshold region is HBCHPT.
Because this approach is the one that was employed in \cite{FBD09} 
and the results presented here are the same as in that reference, 
we restrict ourselves to summarizing the approach
and refer the reader to Ref. \cite{FBD09} for a discussion
of the impact of $D$ waves in the extraction of the $E_{0+}$ multipole, 
of the assessment of the LECs, as well as of the stability of $P$ waves against 
the inclusion of $D$ waves.
In this section we summarize the HBCHPT approach and leave the 
discussion of the
impact of $D$ waves in physical observables to Sec. \ref{sec:results}.

The explicit formulae for the S and P multipoles to one loop
and up to ${\cal O}(q_\pi^4)$ can be found in Refs. \cite{CHPT96,CHPT01}
and constitute the starting point of our analysis.
Due to the order-by-order renormalization process six LECs appear,
and five have been fitted to pion photoproduction data:
$a_1$ and $a_2$ associated with the $E_{0+}$ counterterm $b_p$ associated with the
$P_3$ multipole together with
 $\xi_1$ and $\xi_2$ associated with $P_1$ and 
$P_2$, respectively.
The $c_4$ LEC associated with $P_1$, $P_2$, and $P_3$
has been taken from \cite{MeissnerNPA00}
where it was determined from pion-nucleon scattering inside
the Mandelstam triangle.
Some other parameters appear in the calculation, but these are fixed.
The full list is:
the pion-nucleon coupling constant $g_{\pi N}=13.1$; 
the weak pion decay constant $f_\pi=92.42$ MeV, 
together with the anomalous 
magnetic moments of the proton and neutron,
the nucleon axial charge $g_A$
(which we fix using the Goldberger--Trieman relation $g_A=g_{\pi N}f_\pi/M_p$); and
the masses of the particles.
$D$ waves are included as discussed in the previous section.

In summary, using this approach, in Ref. \cite{FBD09} it was found that,
contrary to what is customarily claimed in the literature, the SP approximation is not sufficient
to obtain a complete description of the differential cross section in the near-threshold region
and to extract the electromagnetic multipoles reliably.
The inclusion of $D$ waves does not affect the extraction of the 
$P$-wave multipoles but makes a significant
difference where 
the $E_{0+}$ extraction is concerned, especially at and above the
unitary cusp.
The absence of $D$ waves in the analysis affects both the $E_{0+}$ multipole
extraction and the determination of the associated LECs.

\subsection{Unitary Fit} \label{sec:unitary}

The Unitary fit is a hybrid approach that employs HBCHPT to compute the $P$ waves
and $E_{0+}$ using Eq. (\ref{eq:Swave})
fitting the parameters $E^{(0)}_{0+}$ and $E^{(1)}_{0+}$ to the experimental data.
These parameters have straightforward relationships to the
LECs $a_1$ and $a_2$ or, more precisely, to their combinations
$a_+=a_1+a_2$ and $a_-=a_1-a_2$.
$E^{(0)}_{0+}$ fixes the value of the multipole at threshold and so does $a_+$ in HBCHPT
\cite{FBD09}.
In the same way, $E^{(1)}_{0+}$ is connected to the combination $a_+ -a_-=2a_2$,
and thus one has a one-to-one relationship between both sets of parameters.

The HBCHPT approach has several shortcomings regarding the $E_{0+}$ multipole.
The first is its slow convergence and the second the lack of unitarity of the amplitude.
In HBCHPT up to one loop and $\mathcal{O} \left( q_\pi^4 \right)$ the value of $\beta = 2.71 \times 10^{-3}/m_{\pi^+}$ is fixed by the imaginary part of $E_{0+}$ ---which is parameter-free--- and 
is not  close to the unitary value of $\beta$ that can be obtained from $\pi N$ scattering and whose value will be discussed in Sec. \ref{sec:resultsA}. The unitarity violation in HBCHPT is due to the truncation at the one loop (single rescattering) level.
It is well known that $E_{0+}$ does not converge very well \cite{book,CHPT91}; however, as will be shown in Sec. \ref{sec:results}, the values for $\text{Re} E_{0+}$  are very close for the HBCHPT and
Unitary fits.

The number of parameters is exactly the same as for the HBCHPT approach ---five, namely, $E^{(0)}_{0+}$ and $E^{(1)}_{0+}$ for the $S$ wave and $b_p$, $\xi_1$ and $\xi_2$ for the $P$ waves, which are the same as those defined in Sec. \ref{sec:hbchpt}.
The $D$ waves are included in the same way as in Sec. \ref{sec:hbchpt}.
Under this approach we have performed two fits, one with $D$ waves and another without them. The results are presented and discussed in Sec. \ref{sec:results}.

\section{Results} \label{sec:results}

\subsection{Multipole Extraction from the Experimental Data.} \label{sec:resultsA}

\begin{figure}
\rotatebox{0}{\scalebox{0.5}[0.5]{\includegraphics{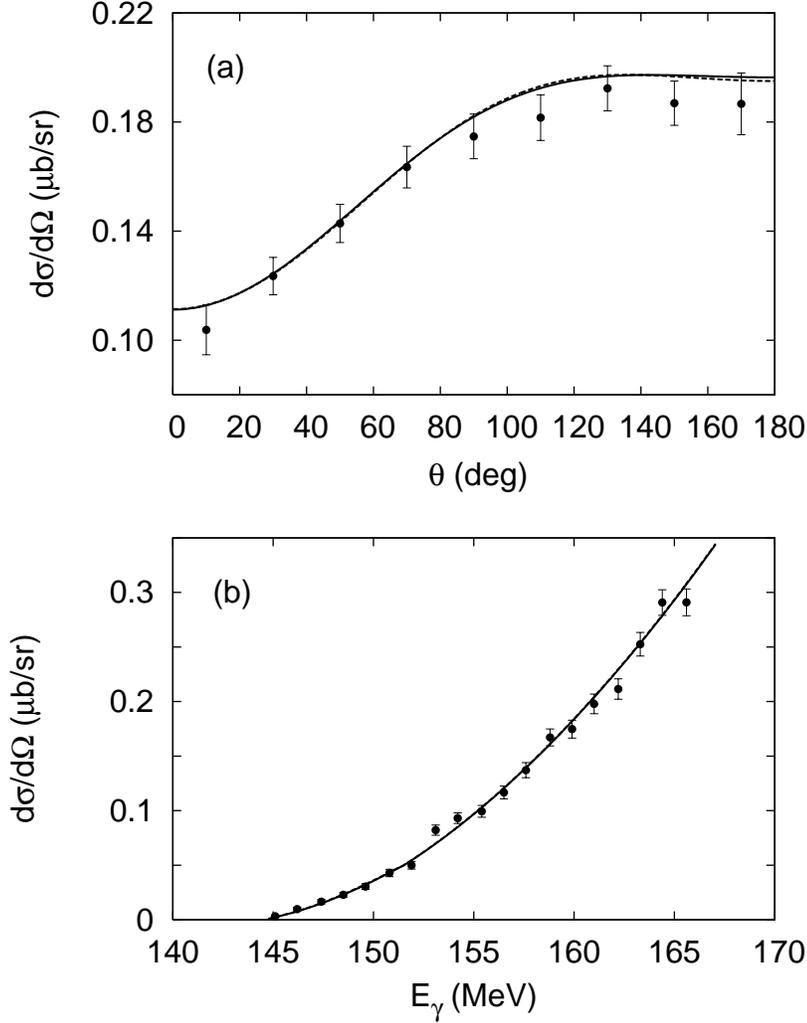}}}
\caption{Differential cross section for a fixed energy of $E_{\gamma}=159.9$ MeV (upper figure) and a fixed angle of  $\theta=90^o$ (lower figure).
Curve conventions: Solid: HBCHPT, SPD fit; Dashed: HBCHPT, SP fit. Both curves almost completely overlap.} \label{fig:Figsig}
\end{figure}

\begin{figure}
\rotatebox{-90}{\scalebox{0.5}[0.5]{\includegraphics{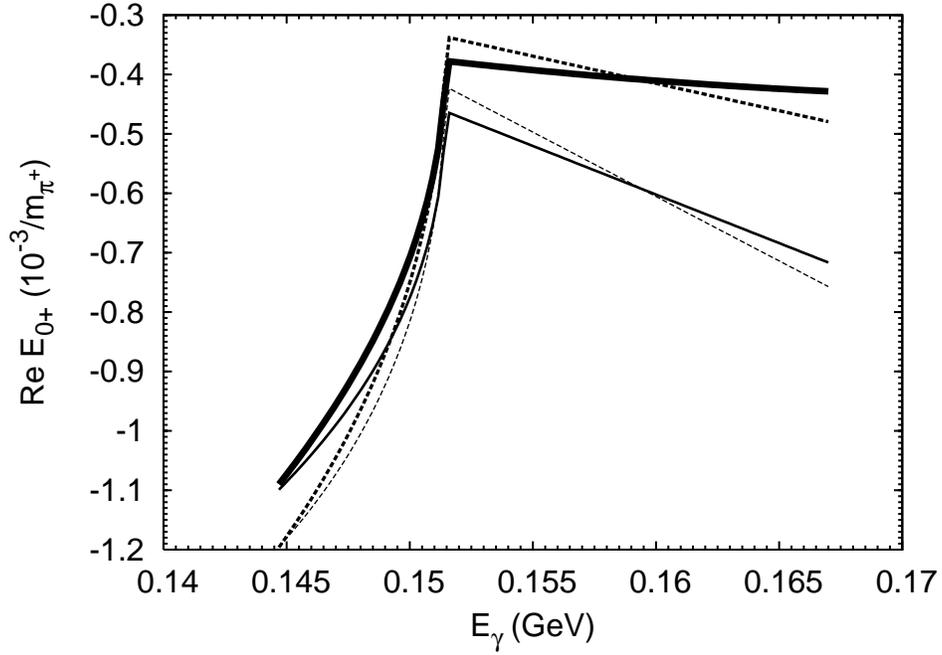}}}
\caption{Extracted $\text{Re} E_{0+}$ multipole.
Curve conventions: Thick Solid: HBCHPT, SPD fit; Thin solid: HBCHPT, SP fit; 
Thick dashed: Unitary, SPD fit; Thin dashed: Unitary, SP fit.} \label{fig:e0+r}
\end{figure}

\begin{figure}
\rotatebox{-90}{\scalebox{0.5}[0.5]{\includegraphics{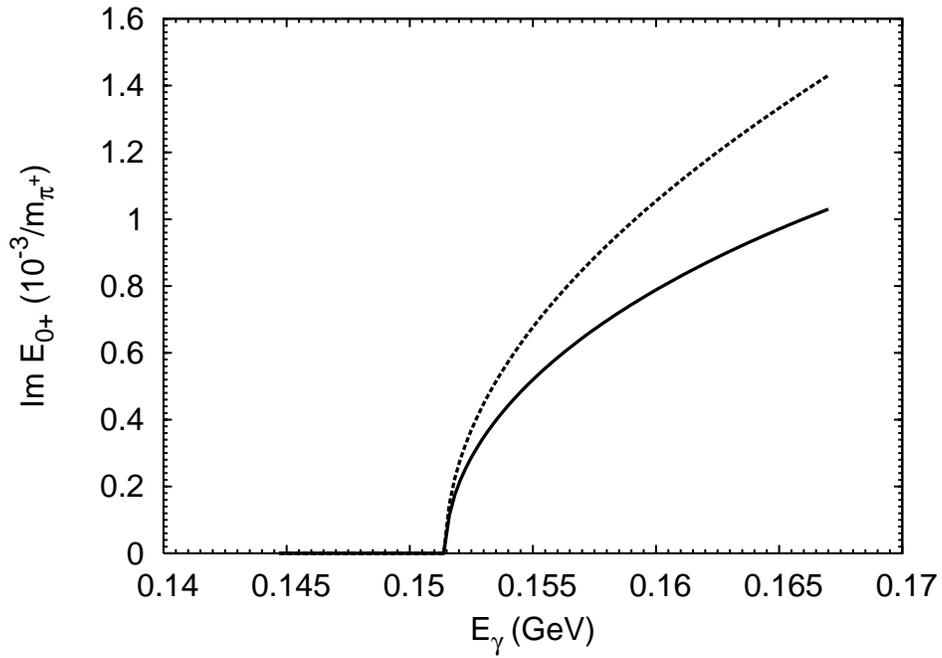}}}
\caption{ $\text{Im} E_{0+}$ multipole. Curve conventions: Solid: HBCHPT (SP and SPD fits);  Dashed: Unitary (SP and SPD fits).} \label{fig:e0+i}
\end{figure}

In Table \ref{tab:summary} we summarize some results for the Unitary
and HBCHPT approaches. 
All of the $\chi^2/$DOF are about the same and are compatible within a 70\% confidence level. 
In Fig. \ref{fig:Figsig} the high quality of the SP \cite{CHPT01,AB-review} 
and SPD fits for HBCHPT can be seen. 
The Unitary fits yield almost identical results.
The $P$ waves are quite alike from one fit to another with differences that lie below the 1.5\% level at threshold. Moreover, the greatest differences are actually at threshold and when the energy increases the differences lie systematically below the 1\% level,
confirming the stability of $P$ waves against the inclusion of $D$ waves in the two approaches.
In Table \ref{tab:summary} we also provide the values at threshold and at the unitary cusp for the real part of  $E_{0+}$. In the next paragraph and with the aid of Figs.  \ref{fig:e0+r} and  \ref{fig:e0+i} we discuss these values.

\begin{table*}
\caption{Summary results for the Unitary and HBCHPT approaches. 
Units of $E_{0+}$ in $10^{-3}/m_{\pi^+}$ and
 $P_i/q_\pi$ in $10^{-3}\text{GeV}^{-1}/m_{\pi^+}$.} \label{tab:summary}
\begin{ruledtabular}

\begin{tabular}{|c|cccc|}

& HBCHPT (SP)   & HBCHPT (SPD) & Unitary (SP)   &  Unitary (SPD) \\
\hline
$\left[ \text{Re} E_{0+} \right]^{\pi^0 \text{thr}}$ & $-1.099$& $-1.090$& $-1.195$& $-1.196$ \\
$\left[ \text{Re} E_{0+} \right]^{\pi^+ \text{thr}} \quad$ &  $-0.478$ & $-0.393$ & $-0.420$ & $-0.336$ \\
$\left[ÊP_1/q_{\pi}\right]^{\pi^0 \text{thr}}$ &73.88&73.99&72.99&73.03 \\
$\left[ÊP_2/q_{\pi}\right]^{\pi^0 \text{thr}}$ &$-$71.84&$-$72.42&$-$71.07&$-$71.48 \\
$\left[ÊP_3/q_{\pi}\right]^{\pi^0 \text{thr}}$ &76.71&76.77&75.98& 76.36\\
$\chi^2/$DOF (minimum) & 1.23 & 1.25 & 1.24 & 1.23 \\
$\chi^2/$DOF(70\% C.L.) &1.27 &1.28 &1.27 &1.26 \\
$\chi^2/$DOF (90\% C.L.) &1.29 &1.30 &1.29 &1.28
\end{tabular}
\end{ruledtabular}
\end{table*}

The main difference between the Unitary and the HBCHPT approaches 
stems from
the associated value of the unitary parameter $\beta$ ---see Eq. (\ref{eq:beta}) ---and how this affects the $S$-wave extraction.  Note that the sign of $\beta$ is observable, not just its magnitude, and agrees with what is expected. The best experimental value of $a(\pi^{-} p \to \pi^{0} n) = -(0.122 \pm 0.002) /m_{\pi^+}$, obtained from the observed width in the $1s$ state of pionic hydrogen \cite{Gotta}, was used. This is in excellent agreement with HBCHPT predictions of $-(0.130 \pm 0.006) /m_{\pi^+}$ \cite{s-pred}. Assuming isospin is a good symmetry, $a(\pi^{+} n \rightarrow \pi^{0} p) = -a(\pi^{-} p \rightarrow \pi^{0} n)$. The latest measurement for 
$E_{0+}({\gamma} p \to \pi^{+} n) = \left( 28.06 \pm 0.27 \pm 0.45 \right)\times 10^{-3}/m_{\pi^+}$ 
\cite{Korkmaz} (where the first uncertainty is statistical and the second is systematic) is in good agreement with the HBCHPT prediction of $\left( 28.2 \pm 0.6 \right)\times 10^{-3}/m_{\pi^+} $ \cite{KR-chiral}. These experimental values and the relationship given above lead to a value of $\beta = \left( 3.43 \pm 0.08 \right) \times 10^{-3}/m_{\pi^+}$.\footnote{Actually, it is energy-dependent, although for the current experimental accuracy it is a very good approximation to assume it is a constant.}

The results obtained in Ref. \cite{FBD09} using HBCHPT regarding the impact of $D$ waves in the 
extraction of the $E_{0+}$ multipole are confirmed using the Unitary approach in 
Sec. \ref{sec:unitary}.
$D$ waves affect the extraction of the $E_{0+}$ multipole and if they are not included a reliable extraction cannot be achieved.
When comparing the $E_{0+}$ extraction using HBCHPT and the Unitary approach it is better to split the analysis above and below the $\pi^+$ threshold. Regarding $\text{Re} E_{0+}$, from Fig. \ref{fig:e0+r} it is evident that above and at the $\pi^+$ threshold
the extractions are very similar if the same number 
of partial waves is used, with only small differences that lie within the uncertainties.
However, if one focuses on the region between the $\pi^0$ and $\pi^+$ thresholds the situation is the opposite. 
As expected, due to the $q_\pi^2$ dependence ---Eq. (\ref{eq:Dwaves}) ---$D$ waves have a negligible impact at $\pi^0$ threshold, as was shown in Ref. \cite{FBD09} for HBCHPT.  As shown in Fig. \ref{fig:e0+i}, there is a sizeable difference between the two approaches 
regarding the imaginary part of $E_{0+}$, with the Unitary approach prediction being larger than that for the HBCHPT approach due to the larger value of $\beta$. The piece of $E_{0+}$ that contributes to the imaginary part above the unitary cusp contributes to the real part below it. This means that this 
contribution in the real part of $E_{0+}$ near threshold is larger for the Unitary approach than for HBCHPT (this is the effect that we find at threshold) and makes the value of $\left[ \text{Re} E_{0+} \right]^{\pi^0 \text{thr}}$ in Table \ref{tab:summary} larger in absolute value for the Unitary approach than for HBCHPT. Hence, an accurate determination of the actual value of $E_{0+}$ at threshold would help one
to constrain the imaginary part of the multipole and vice versa.

\begin{figure}
\rotatebox{-90}{\scalebox{0.5}[0.5]{\includegraphics{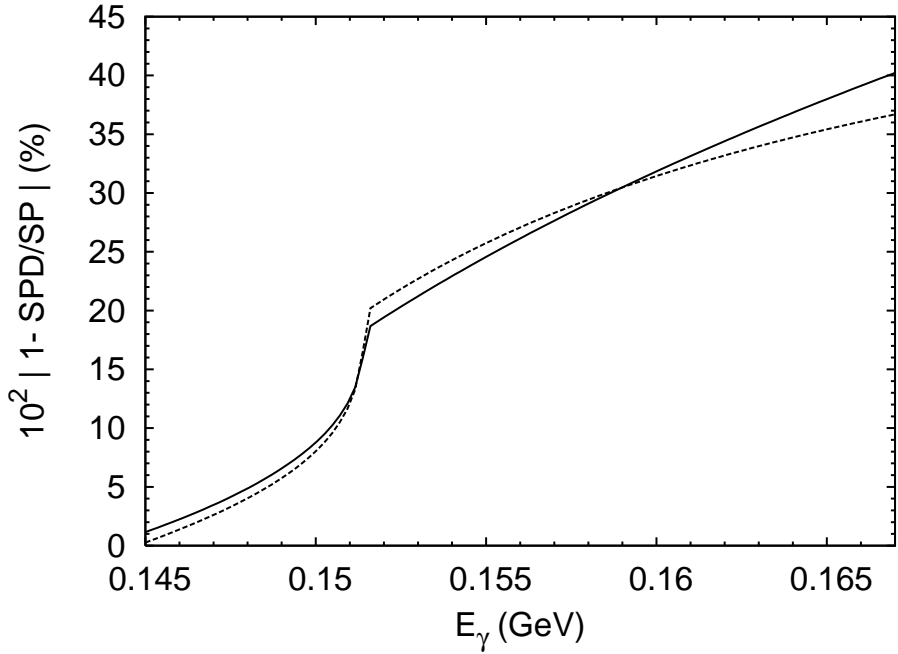}}}
\caption{Extracted $\text{Re} E_{0+}$ multipole ratio between the SP and SPD fits. 
Curve conventions: Solid: HBCHPT; Dashed: Unitary.} \label{fig:SPvsSPD}
\end{figure}

In Fig. \ref{fig:SPvsSPD} we focus on the effect of $D$ waves, comparing the SP and the SPD fits for both HBCHPT and Unitary approach by computing the ratio for the real part of $E_{0+}$.
The impact of $D$ waves is astonishing, changing
the value of the multipole by almost 20\% at the unitary cusp and reaching a 35\% change at 165 MeV.

In Ref. \cite{VKM2005} the $E_0+$ multipole [employing the phenomenological formula
in Eq. (\ref{eq:Smeissner})] was obtained fitting 
the subtraction constants to the dispersion analysis of Ref. \cite{Pasquini}
to compute the Fubini--Furlan--Rossetti (FFR) sum rule \cite{Fubini}.
This approach has the shortcoming that it relies on the extraction made through a dispersion analysis
that is model dependent because it relies on MAID03 \cite{MAID} to obtain the $A_1$ amplitude that enters in the FFR sum rule. It does have the advantages that HBCHPT
has better convergence inside the Mandelstam triangle and that all partial waves are incorporated
in the dispersion analysis.

\subsection{Impact of $D$ waves in Observables}  \label{sec:resultsB}

\begin{figure}
\rotatebox{0}{\scalebox{0.5}[0.5]{\includegraphics{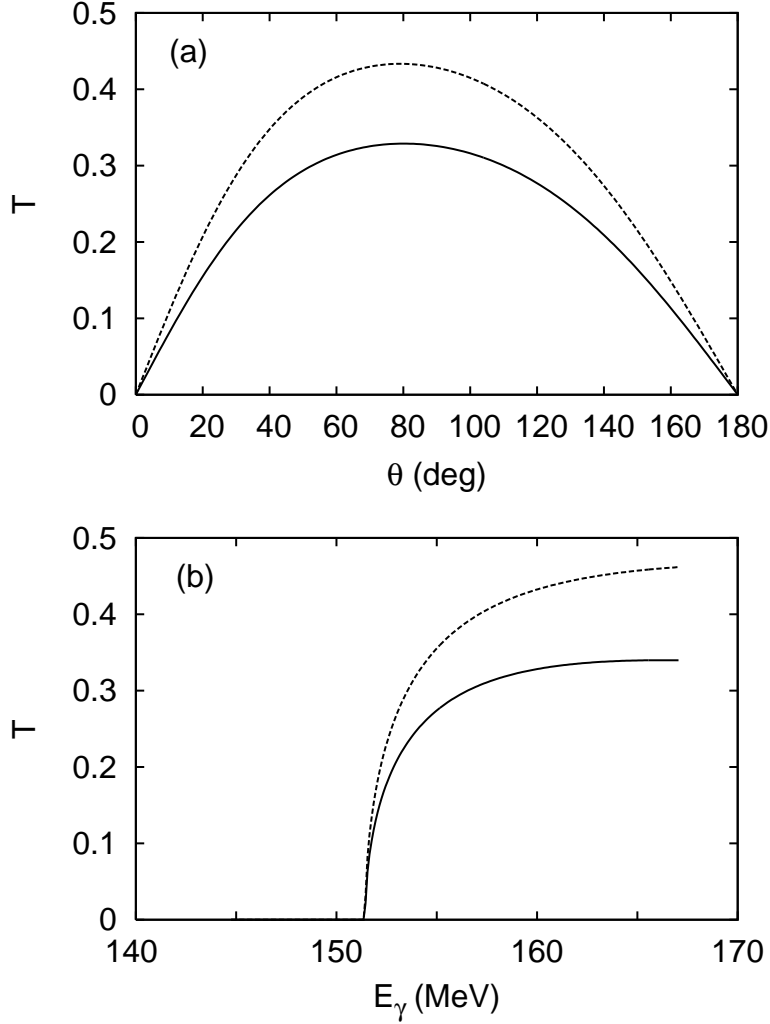}}}
\caption{$T$ asymmetry for a fixed energy of $E_{\gamma}=159.5$ MeV (upper figure) and a fixed angle of  $\theta=90^o$ (lower figure).
Curve conventions: Solid: HBCHPT, SPD fit; Dashed: Unitary, SPD fit.} \label{fig:Tasym}
\end{figure}

\begin{figure}
\rotatebox{0}{\scalebox{0.5}[0.5]{\includegraphics{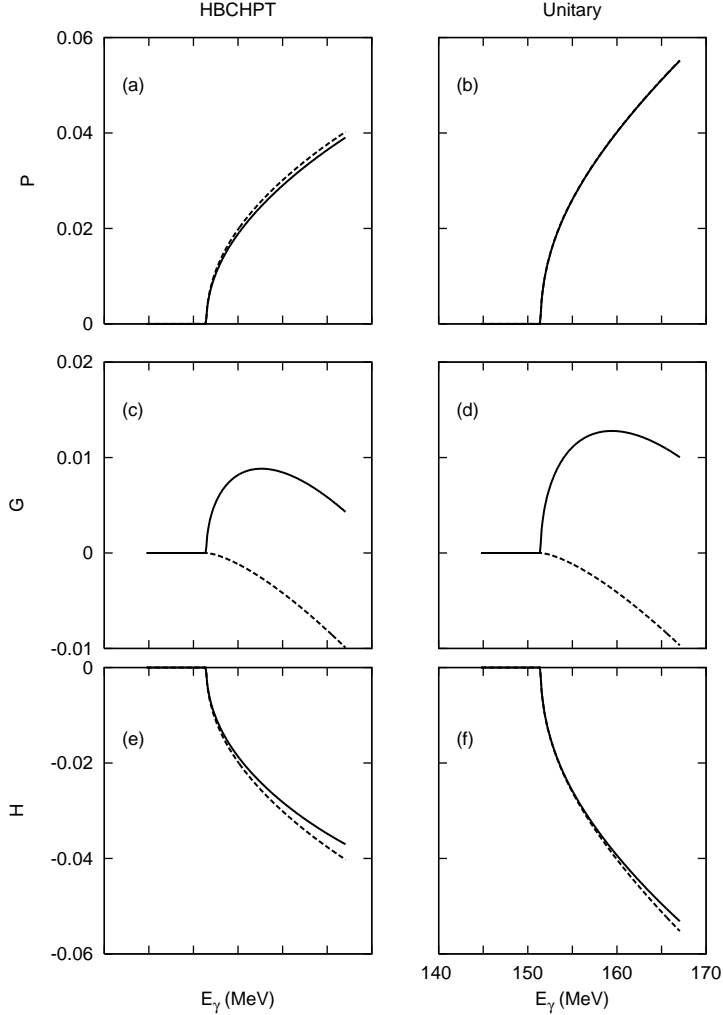}}}
\caption{TRO observables $P$, $G$, and $H$ at outgoing pion angle $\theta_{cm}=90^o$ for SP and SPD fits using HBCHPT and the Unitary approaches. Curve conventions: Solid: SPD fit; Dashed: SP fit.
Data are from \cite{Schmidt}. In Fig. (b) both curves almost completely overlap.} \label{fig:Pol_90PGH}
\end{figure}

If the physical observables (differential cross section and asymmetries) are computed using either HBCHPT or the Unitary approaches, we do not find meaningful differences except for the case of TRO observables ($T$, $P$, $G$, and $H$; see, Figs. \ref{fig:Tasym} and \ref{fig:Pol_90PGH}) 
and, among them, the only one that is measurable 
in the near-threshold region
is the $T$ asymmetry\footnote{Do not confuse the $T$ asymmetry symbol with the $T_i$ which stand for the partial wave expansion of the differential cross section. The latter has a subscript.}
defined as the difference between the cross sections for a target polarized up and down along the axis perpendicular to the scattering plane.
In Fig. \ref{fig:Tasym} we compare the results obtained for $T$ 
using the Unitary and the HBCHPT approaches including $D$ waves. 
These differences make the $T$ asymmetry an excellent observable
to test unitarity and to extract the imaginary part of $E_{0+}$,
because the latter is largely responsible of the difference.
At present there are proposals 
to measure this asymmetry both at HI$\gamma$S \cite{AB-review} and MAMI \cite{MAMI-Bernstein}.

Henceforth we will focus the discussion 
on the effects of $D$ waves on observables using the HBCHPT calculation.
This particular choice does not affect either the discussion or the conclusions.

In Fig. \ref{fig:Pol_90PGH} we present the comparison
between the HBCHPT and Unitary approaches, both SP and SPD fits, for the TRO
observables:
recoil nucleon polarization asymmetry $P$,
and double beam target polarized asymmetries $G$ and $H$ 
all as functions of the photon energy and for a fixed angle of $90^o$
for HBCHPT and the Unitary approach. 
The results are similar at other angles.
The definition  can be found in Ref. \cite{BDS75}
or in Ref. \cite{AB-review};
in this article we follow the sign conventions of Ref. \cite{CT97}.

\begin{figure}
\rotatebox{0}{\scalebox{0.5}[0.5]{\includegraphics{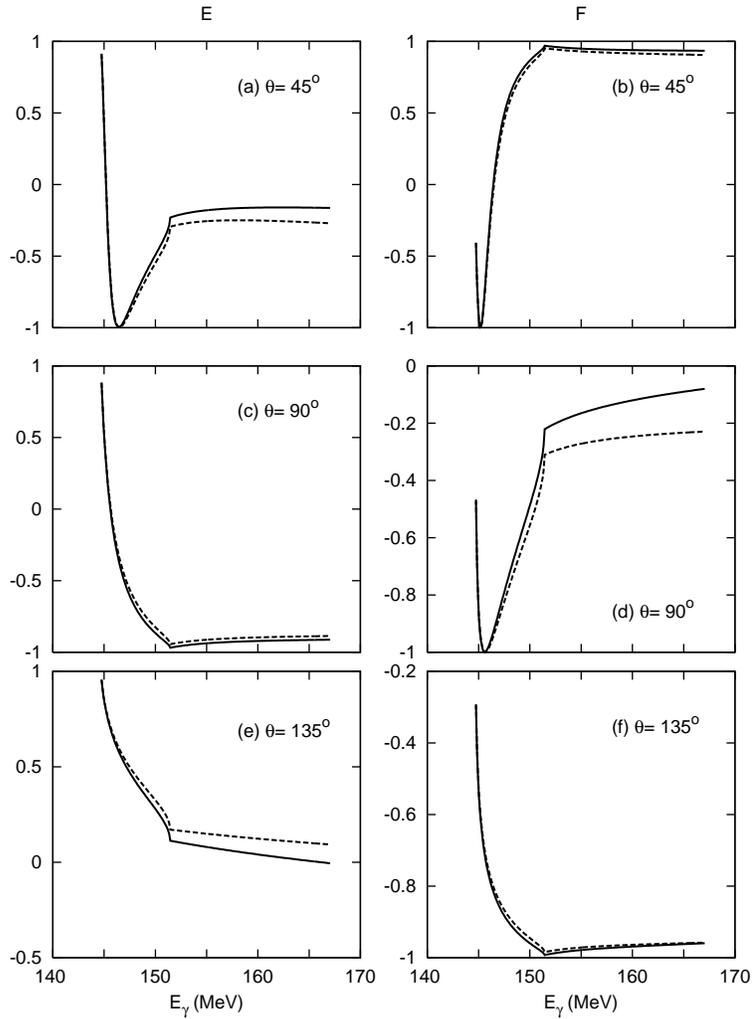}}}
\caption{$E$ and $F$ asymmetries for SP and SPD fits using HBCHPT. Curve conventions: Solid: HBCHPT, SPD fit; Dashed: HBCHPT, SP fit.} \label{fig:EF_panel}
\end{figure}

The effect of $D$ waves in the $G$ asymmetry is striking, 
but unmeasurable at present, since it is so small. 
The same can be said about
$P$ and $H$: the necessary accuracy to measure 
the effect is at present 
beyond the current state-of-the-art.
Differences between the Unitary and HBCHPT approaches are also unmeasurable at present.
The most promising observables are the $E$ and $F$ asymmetries and
in Fig. \ref{fig:EF_panel} we present them 
in terms of their energy dependence, which highlights the $D$-wave effect
for three different pion production angles: $45^o$, $90^o$, and $135^o$.
When the SP and SPD fits yield different results, the predictions obtained using the
HBCHPT and Unitary approaches are found to be significantly closer together than
these differences. Therefore, in the remainder of the article we restrict the discussion to the
HBCHPT fits.

The combined measurement of the $E$ and $F$ asymmetries provides
a good candidate for determining the $D$ waves.
The reason is that both are sizeable and behave in opposite ways for the $D$-wave inclusion: for those angles where $D$ waves do not affect the $E$ asymmetry they do impact the $F$ asymmetry and vice versa (see Fig. \ref{fig:EF_panel}). More explicitly, at $90^o$ the $F$ asymmetry yields sizeable differences for the SP and SPD predictions, while not at $45^o$ and $135^o$.
The $E$ asymmetry happens to be just the opposite.
Hence, an analysis of both asymmetries using the SP and SPD approximations should be
better at showing up the effects of $D$ waves.

\begin{figure}
\rotatebox{0}{\scalebox{0.5}[0.5]{\includegraphics{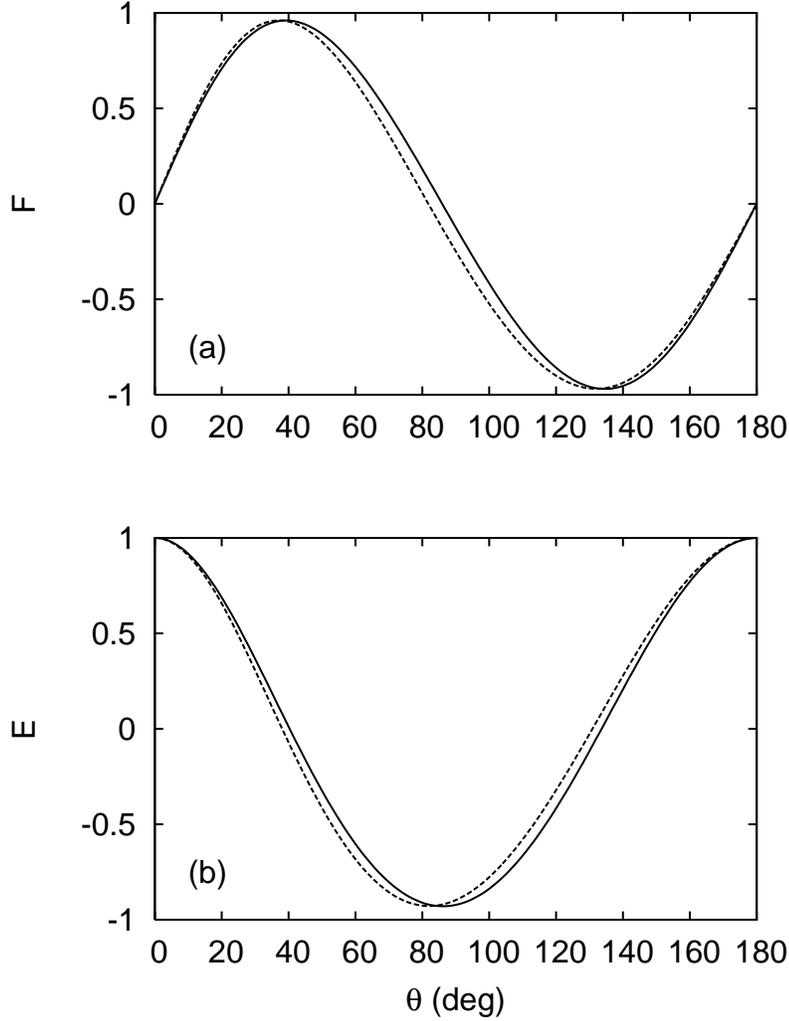}}}
\caption{Angular dependence for the  $E$ and $F$ asymmetries at $E_{\gamma}=159.5$ MeV. Curve conventions: Solid: HBCHPT, SPD fit; Dashed: HBCHPT, SP fit.} \label{fig:FEasym}
\end{figure}

\begin{figure}
\rotatebox{-90}{\scalebox{0.5}[0.5]{\includegraphics{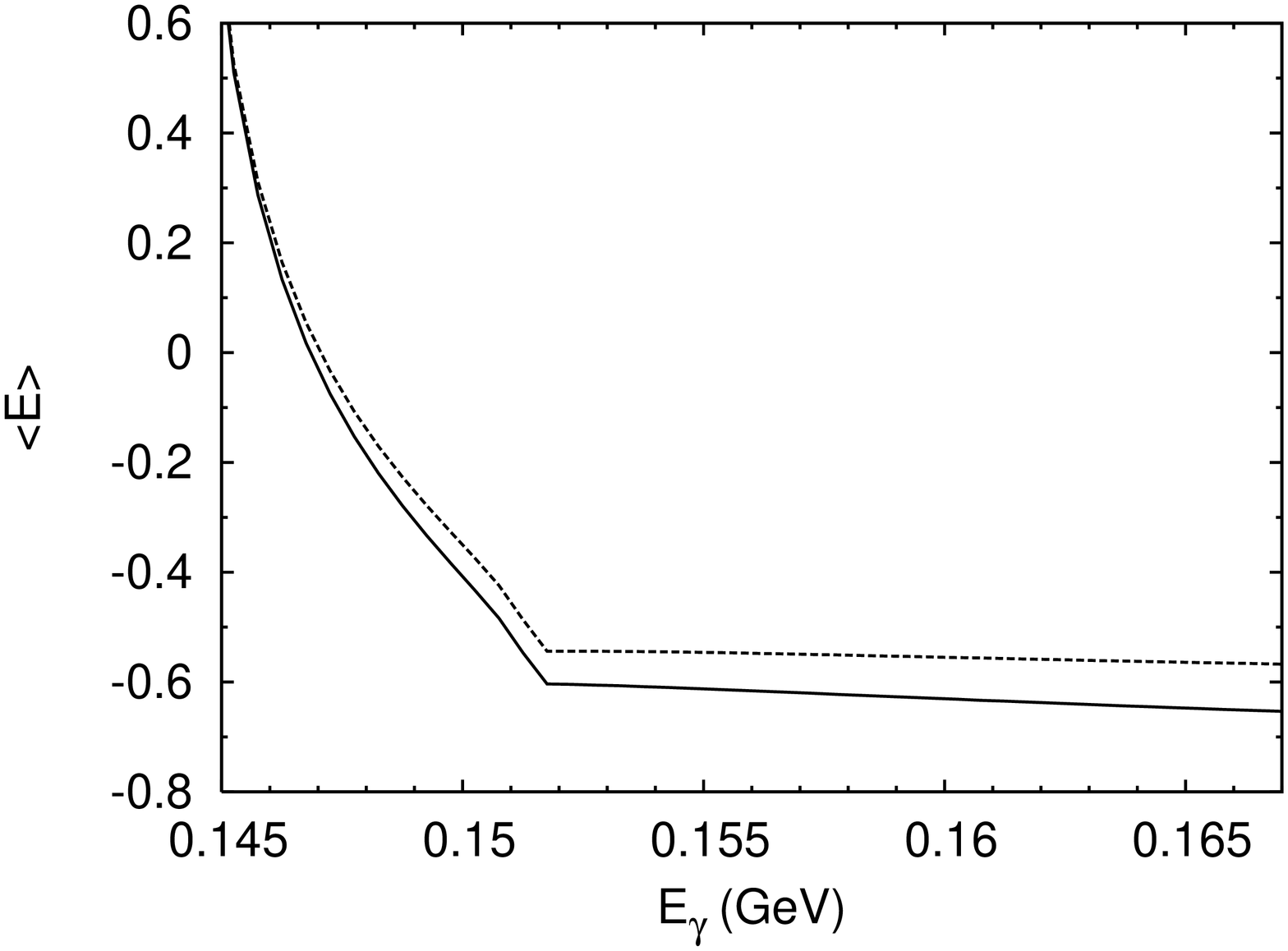}}}
\caption{$E$ asymmetry depending on energy and integrated
in the angular range from $90^o$ to $140^o$.
Curve conventions: Solid HBCHPT, SPD fit; Dashed: HBCHPT, SP fit. 
Errors are computed at the 70\% confidence level.} \label{fig:Easym}
\end{figure}

\begin{figure}
\rotatebox{-90}{\scalebox{0.5}[0.5]{\includegraphics{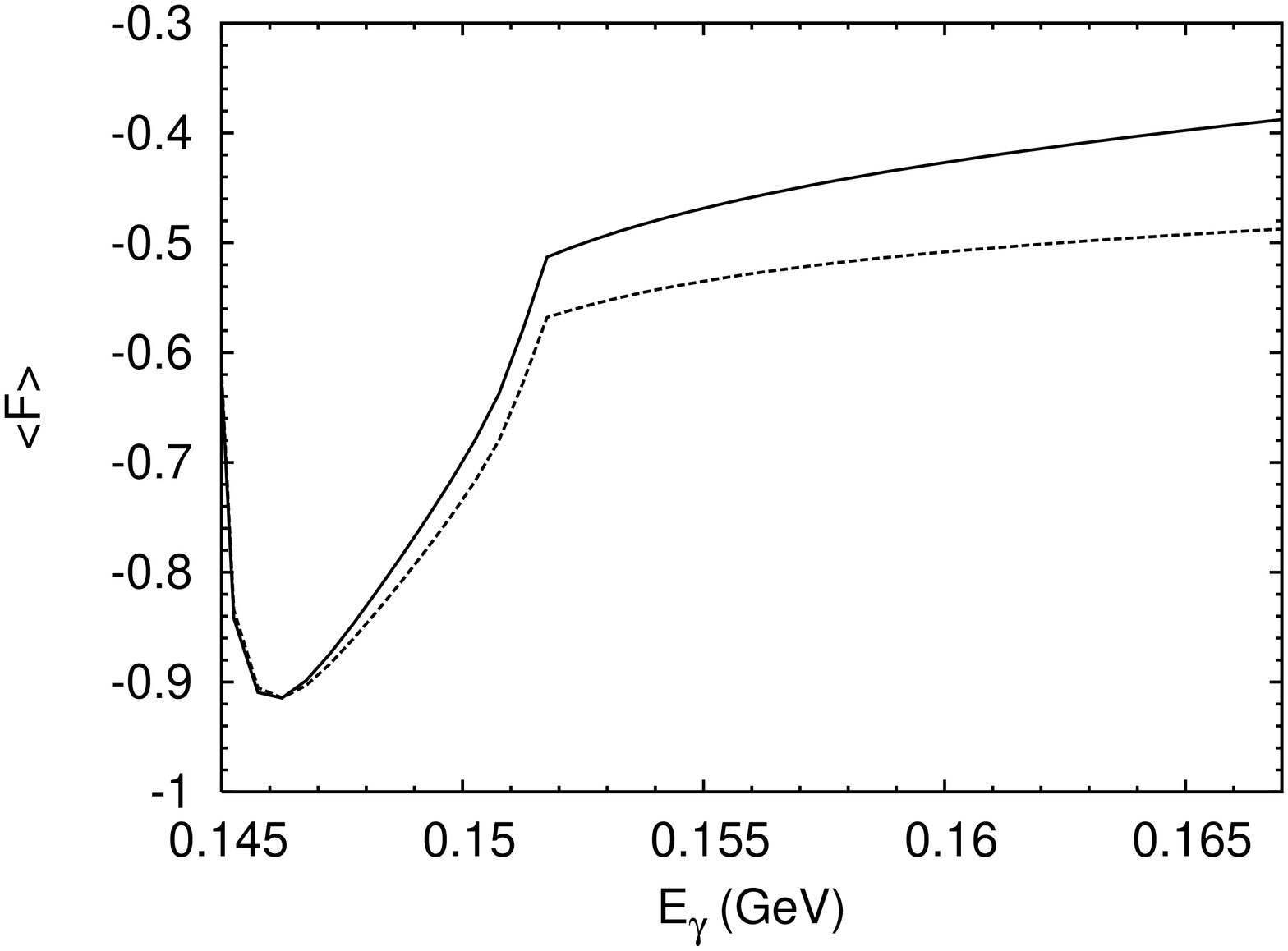}}}
\caption{$F$ asymmetry depending on energy and integrated
in the angular range from $70^o$ to $140^o$. 
Curve conventions: Solid: HBCHPT, SPD fit; Dashed: HBCHPT, SP fit. 
Errors are computed at the 70\% confidence level.} \label{fig:Fasym}
\end{figure}

In Fig. \ref{fig:FEasym} we display the angular dependence of $E$ and $F$
for an intermediate energy of
$E_{\gamma} = 159.5$ MeV.\footnote{We have chosen this energy both because it is intermediate and because it is the only one for which there are available $\Sigma$ data.}
The differences between results for the SP and SPD fits are not very apparent,
but become more so when the observable is integrated over some convenient angular range 
($\theta_i, \theta_f$); for instance, for the $E$ asymmetry one might use
\begin{equation}
<E>=\frac{ \int^{\theta_f}_{\theta_i}  W_E \left( s, \theta \right) \sin \theta d\theta}{ \int^{\theta_f}_{\theta_i}  W_T \left( s, \theta \right) \sin \theta d\theta} \: , \label{eq:<E>}
\end{equation}
where $W_T$ and $W_{E}$ are defined by Eqs. (\ref{eq:wt}) and (\ref{eq:we}).
This allows one to take advantage of
cancellations in the asymmetry, thereby enhancing the features in which
we are interested, and allows one to optimize the value of the figure of merit (FOM) ---for the E asymmetry the $FOM=\sigma_T <E>^2$ --- and hence minimize the statistical uncertainties. The observable $<F>$ can be defined in a similar way for the $F$ asymmetry by replacing $W_E$ in 
Eq. (\ref{eq:<E>}) with $W_{F}$ from Eq. (\ref{eq:wf}).
In Figs. \ref{fig:Easym} and \ref{fig:Fasym} we provide $<E>$ and $<F>$ integrated 
over the appropriate angular ranges for the Crystal Ball at MAMI that maximize 
their FOMs \cite{MAMI-Bernstein}.

The main contribution to the differences between the SP and SPD fits
in $<E>$ are due first,  to the differences in the $E_{0+}$ multipole and, second, to the interference of the $M_{1+}$ multipole with
 $E_{2-}$. Again, the interference of $D$ waves with $M_{1+}$ constitutes the leading $D$-wave correction.
However, the fact that the biggest deviation is due to the modification of the $E_{0+}$ multipole
proves that $D$ waves are effectively entangled with the $S$ wave and that the extraction of any of them must rely on the knowledge of the others.

These asymmetries are large and actually would allow one to find a trace of $D$ waves.
However, we believe that the combined measurement of both would improve the analysis
beyond doubt. Nevertheless, the angular distribution is also of great interest in order to obtain the coefficients $F_i$ and $E_i$ in Eqs. (\ref{eq:wf}) and (\ref{eq:we}).
There are plans to measure $F$ at HI$\gamma$S \cite{AB-review} and MAMI \cite{MAMI-Bernstein}.

\begin{figure}
\rotatebox{0}{\scalebox{0.5}[0.5]{\includegraphics{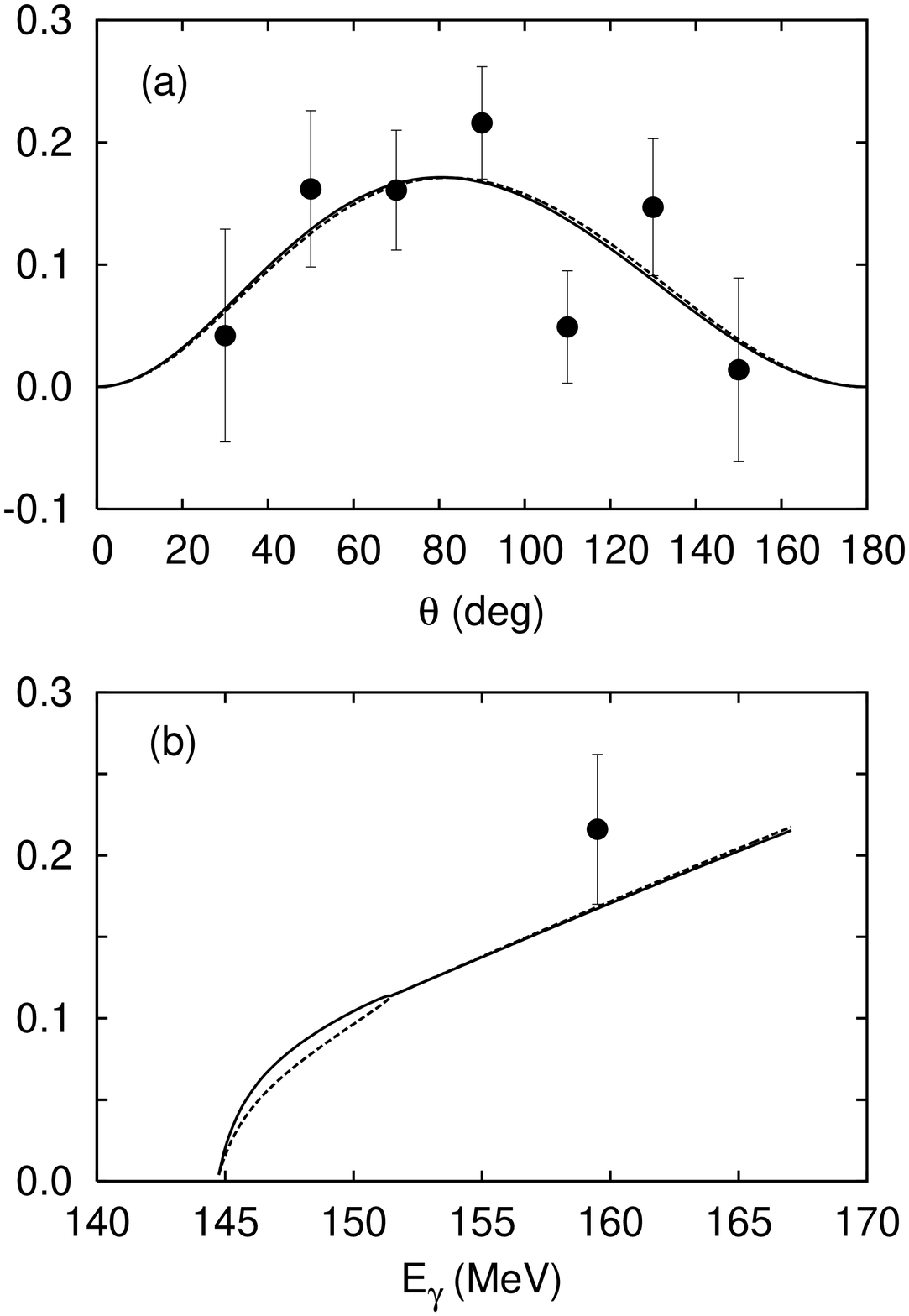}}}
\caption{$\Sigma$ asymmetry for a fixed energy of $E_{\gamma}=159.5$ MeV (upper figure)
and a fixed angle of  $\theta=90^o$ (lower figure).
Curve conventions: Solid HBCHPT, SPD fit; Dashed: HBCHPT, SP fit.} \label{fig:Sigma1595}
\end{figure}

At first thought, the photon beam asymmetry $\Sigma$ seems to be a promising
observable to pin down $D$ waves if enough accuracy is achieved.
If we consider only $S$ and $P$ waves, this asymmetry can be written:
\begin{equation}
- W_T \Sigma = S_0 \sin^2 \theta \: , \label{eq:TT0}
\end{equation}
where $S_0$ can be read from Table \ref{tab:TT0}.
The inclusion of $D$ waves modifies $S_0$ and adds two new terms 
$S_1 \mathcal{P}_1 \left( \theta \right)\sin^2\theta$ 
and $S_2 \mathcal{P}_2 \left( \theta \right)\sin^2\theta$ to the right-hand side of 
Eq. (\ref{eq:TT0}). The leading term would be $S_1$;
however, this term seems to be very small and difficult to isolate,
as we show in Fig. \ref{fig:Sigma1595} where we compare the
SP and SPD fits with experimental data \cite{Schmidt}.
Moreover, the $D$-wave contribution could be easily hidden by the fitting, embedding the $D$ waves in the effective parameters of the multipoles. In order to avoid this situation, some other observables should be measured in order to find inconsistencies in the SP fits that can only be solved by adding new terms ($D$ waves).

For this reason, one should also consider some other convenient observables. For instance, if the $E$ asymmetry is measured together with the differential cross section and enough angular points are acquired,
it is possible to extract the values of $T_i$ and $E_i$ at a certain energy. If we compare the tables for $T_1$ and $E_1$ (Tables \ref{tab:T1} and \ref{tab:E1}) we notice that both have the same coefficients for the $S$-$P$ interference, and hence any difference between them is due to higher partial waves ($D$ waves). Moreover, if we assume $E_{1+}\simeq 0$ (which is actually a sensible assumption), it is straightforward to obtain
\begin{equation}
\begin{split}
\mathfrak{B}\equiv&T_1 - E_1 \\
\simeq &\frac{18}{5} \text{Re}  \{  M_{1+}^*\left[  4 \left( M_{2+} -E_{2+}\right)+M_{2-} +E_{2-} \right] \} \:,
\end{split} \label{eq:T1E1}
\end{equation}
whose measurement provides a perfect test of the contribution of $D$ waves because the
$P$ wave $M_{1+}$ is large and well known.

\begin{figure}
\rotatebox{-90}{\scalebox{0.5}[0.5]{\includegraphics{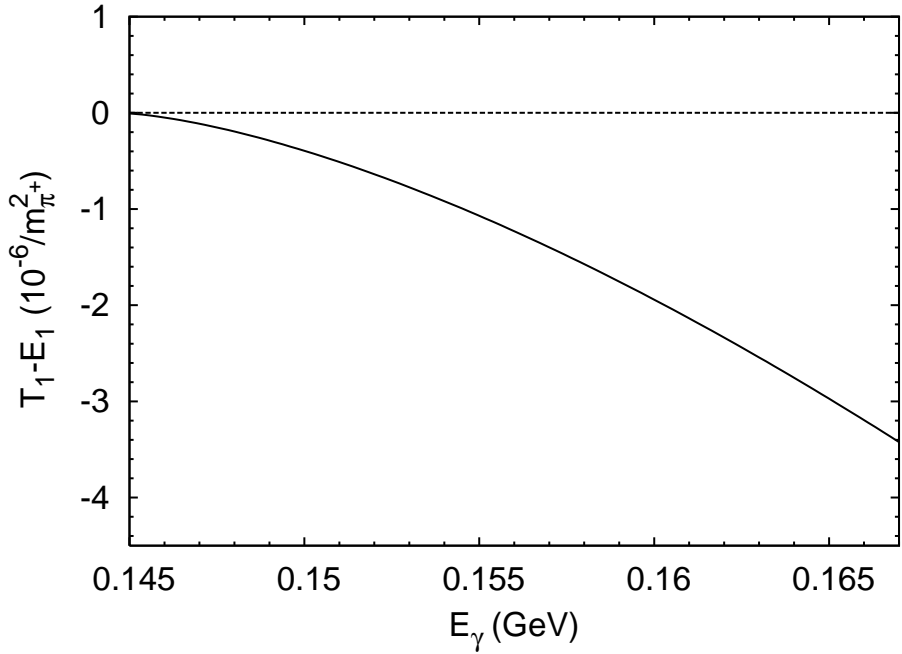}}}
\caption{$\mathfrak{B}\equiv T_1-E_1$ for HBCHPT SPD fit (solid) and HBCHPT SP (dashed). The SP calculation yields zero and is shown for comparison purpouse.} \label{fig:T1-E1}
\end{figure}

Both pieces of information can be measured and one can obtain an unmistakable trace of $D$ waves in low-energy pion photoproduction due to the fact that under the SP approximation this difference is exactly zero. Hence, any deviation from zero will be due to the interference of $P$ and $D$ waves. Moreover, as long as $M_{1+}$ is the dominant $P$-wave contribution (and well known) and $E_{1+}$ 
is negligible, the $D$-wave combination 
$4 \left( M_{2+} -E_{2+}\right)+M_{2-} +E_{2-}$ can be isolated,
providing a test of its presence or absence. 
In Fig. \ref{fig:T1-E1} we show our prediction for $\mathfrak{B}$, 
whose magnitude is possibly measurable with the current state-of-the-art facilities.
This would also provide insight on the $D$ waves, 
testing how accurate is the Born+vector-meson approach employed in this article.

\begin{figure}
\rotatebox{-90}{\scalebox{0.5}[0.5]{\includegraphics{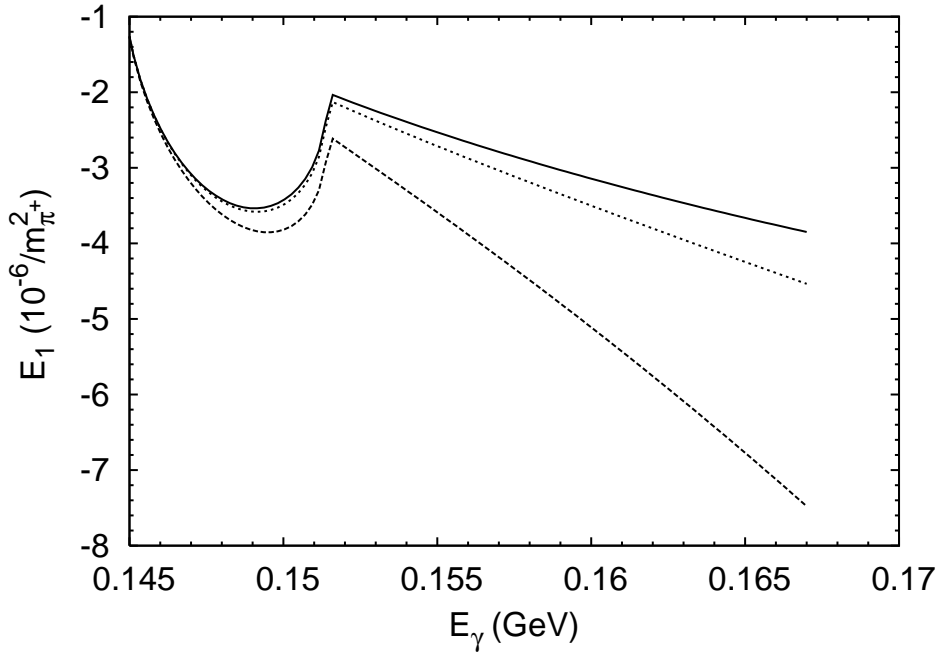}}}
\caption{$E_1$ for fits HBCHPT SPD (solid), HBCHPT SP (dashed), and HBCHPT SPD (dotted) with the $D$ waves removed.} \label{fig:E1}
\end{figure}

It is worth noticing an important difference between this observable and the measurement of $E$ and $F$ in the search of $D$ waves. This combination is truly a measurement of $D$ waves, while in $E$ and $F$ what is involved is the impact of $D$ waves on the extraction of the $S$-wave. Actually, the main contribution to the difference between the SP and SPD fits is not the addition of $D$ waves, it is the modification of the $S$ wave due to the addition of $D$ waves and the subsequent impact of this difference on the observables. Figure \ref{fig:E1} illustrates this statement. We display the results from three calculations for the $E_1$ observable. The solid line is the HBCHPT SPD fit and the dotted one the HBCHPT SP one. The difference between the two is apparent.
The third curve is the HBCHPT SPD fit without the $D$ waves. As long as $P$ waves are almost equal for the two calculations, the difference between the dotted and dashed curves is due to the difference between the $S$ waves, and the difference between the solid and the dotted curves is the true contribution of the $D$ waves, mainly through their interference with $P$ waves.

Another good candidate for extracting the $E_{0+}$ multipole is 
\begin{equation}
\begin{split}
\mathfrak{D} \equiv T_0+2E_0&= 3|E_{0+}|^2 + 3|M_{1-}|^2  \\
&+ 12 |E_{1+}|^2 + 12 \text{Re} \{  E^*_{1+} M_{1+}\}+ \delta D \: , \label{eq:donnellyasym}
\end{split}
\end{equation}
where $\delta D$ stands for the $D$-wave content, namely $D$ waves interfering with $D$ waves, which guarantees a very small contribution to the observable. As long as the $P$ waves are well known from HBCHPT and $E_{1+}$ is very small, then
we can extract the value of $|E_{0+}|^2$ if we know the $P$-wave $M_{1-}$.
If we approximate  $\mathfrak{D} \simeq 3|E_{0+}|^2 + 3|M_{1-}|^2$, the error is less than a 
6\%.
Combined with other measurements,  $\mathfrak{D}$ would help to pin down the $E_{0+}$ multipole.

Actually, the particular choice $\mathfrak{D}$ can be exploited in a different way.
Experimentally, asymmetries can be measured more accurately than cross sections, and the quantities
$E_0$, $E_1$, $F_0$ and $F_1$ can be measured with relatively small systematic
errors.
Moreover, we can define
\begin{equation}
W_{T}=T_0\left[Ê1+ \tilde{T}_1 \mathcal{P}_1\left( \theta \right)  
+ \tilde{T}_2 \mathcal{P}_2\left( \theta \right) + \dots \right]
\end{equation}
with $\tilde{T_1}\equiv T_1/T_0$ and $\tilde{T_2}\equiv T_2/T_0$ 
that can both also be measured very accurately.
Hence, the key quantity to improve is $T_0$, which is the total cross section up to a kinematical factor
\begin{equation}
\sigma \left( s \right) = 2\pi \frac{q_\pi}{k_\gamma} T_0 \: .
\end{equation}

From accurate measurements of the asymmetries ($T$, $F$ and $E$) it is possible to extract the multipoles and use this knowledge to isolate $T_0$ in Eq. (\ref{eq:donnellyasym}), providing
a procedure to determine the cross section more accurately.
These are a few of the analyses
that a deep knowledge of the structure of the observables 
in terms of the multipoles allows one to undertake using
the tables in Appendix \ref{sec:apendiceA}.

\subsection{Contribution of Higher Partial Waves to the Observables} \label{sec:resultsC}

At this point a fair question to ask is why should we stop at $D$ waves and not include $F$ waves in the analysis. Within the current experimental accuracy (and probably for years to come) the answer is
that $F$ waves do not play a role and can be dismissed.
The reason to make this statement relies on the structure of the observables presented in Appendix \ref{sec:apendiceA}. Looking at the expansions in Eqs. (\ref{eq:wt}), (\ref{eq:wtt}), (\ref{eq:we}) and (\ref{eq:wf}), coefficients such as $T_3$, $T_4$, $S_2$, $E_3$, $E_4$, $F_2$, and $F_3$  are negligible experimentally. Coefficients such as $T_0$, $T_2$, $S_0$, $E_0$, and $F_1$ are $|M_{1+}|^2$ and $|M_{1-}|^2$ dominated, and so if $F$ waves would have impact the only observables that remain are those where $D$ waves already make an impact, namely $T_1$, $S_1$, $E_1$, and $F_0$.
At this point we remind the reader that the enhancement of $D$ waves is due to their interference with the large $P$-wave multipoles, $M_{1+}$ (primarily)
and $M_{1-}$, and therefore any non-negligible 
$F$ wave contribution should be related to the same kind of interference.
Due to symmetry reasons, in $T_1$, $E_1$, and $F_0$ 
$F$ waves interfere only with $D$ and $G$ waves. 
The argument is as follows: first, for $T_1$, $E_1$ and $F_0$ 
only partial waves with opposite parity can interfere
(see Tables \ref{tab:T1}, \ref{tab:E1} and \ref{tab:F0}); 
second, these partial waves can differ only by one unit of angular momentum.
$S_1$  is the only coefficient where $F$ waves interfere with $P$ waves, but we have found that 
$D$-wave/$P$-wave interferences for this observable are small and hence we expect $F$-wave/$P$-wave interferences to be even smaller. So $F$ waves are highly suppressed in the observables and can be safely neglected in the near-threshold region
within the current experimental state of the art.
Clearly our expectation is that $G$ waves and higher play
an even more minor role in the near-threshold region.

\section{Summary and conclusions}

In Ref. \cite{FBD09} neutral pion photoproduction from the proton was studied
in the region from threshold up to $167$ MeV using HBCHPT.
The $E_{0+}$ multipole and $D$ waves 
are entangled in such a way that the accurate
determination of the former 
(including the extracted values of the LECs $a_1$ and $a_2$) 
relies on the proper inclusion of $D$ waves.
In this article we complement the analysis in Ref. \cite{FBD09} by employing a phenomenological approach that needs more experimental input to draw reliable conclusions
and a Unitary approach
that uses HBCHPT for $P$ waves and a unitary prescription for the $E_{0+}$ multipole.

We have found no impact of $D$ waves at $\pi^0$ threshold on the $S$ wave,
although a noticeable difference is found between the Unitary and HBCHPT approaches 
due to the difference in the value of $\beta$.
We have also considered partial waves higher than $L=2$
and have found that they are not important and can be neglected.
We have found that $P$ waves are stable  at the 1\% level
against the inclusion of $D$ waves or the use of a different prescription
for the $E_{0+} $ multipole.

The major difference between the SP and SPD calculations
of the observables does not rely on the direct
contribution of $D$ waves, but on the modification of the $E_{0+}$ multipole, 
which is misleadingly extracted from data
if only $S$ and $P$ waves are included.
Although $D$ waves are typically weak, the modification of the $S$ wave is sufficiently large
to alter the prediction for the 
double-polarization observables $E$ and $F$ enough to
to be able to distinguish between approaches with and without $D$ waves,
leading to measurable effects that will allow one to  pin down $D$ waves experimentally.
In the next generation of  photo-pion experiments \cite{AB-review,MAMI-Bernstein,Hornidge} these could be measured at MAMI and HI$\gamma$S. Specifically, it has been demonstrated that the statistical accuracy of the recently proposed polarization experiment to determine $F$ at MAMI will be sufficient to unveil the importance of the $D$-wave contribution. In this article we have gone further and shown also that the $E$ observable is sensitive to the $D$-wave contribution and that the combination $T_{1} -E_{1}$ will provide a direct quantitative measure of the $D$-wave effect (Fig. \ref{fig:T1-E1}). We thus expect in the next few years a direct experimental test of the ideas put forward in this article. 

We also explored the influence of $D$ waves in $\pi N$ scattering and found no sizeable impact on the
partial-wave extraction. In addition, the treatment of the Coulomb interaction poses uncertainties that are larger than the estimated $D$-wave effects.

In conclusion, $D$ waves cannot be dismissed in the analysis of low-energy neutral
pion photoproduction from the proton
due to the soft nature of the $S$ wave
that is direct consequence of chiral symmetry and the Nambu--Goldstone nature of the pion.

\begin{acknowledgments}
This research was supported in part (CF-R) by 
 \textquotedblleft Programa Nacional de Movilidad de Recursos Humanos del Plan Nacional I+D+I 2008-2011\textquotedblright
of Ministerio de Ciencia e Innovaci\'on (Spain).
This work was also supported in part (AMB and TWD)
by the U.S. Department of Energy under contract No. DE-FG02-94ER40818. 
\end{acknowledgments}

\appendix

\section{Physical Observables in Terms of the Electromagnetic Responses: Multipolar Tables up to $D$ Waves} \label{sec:apendiceA}

The differential cross section and asymmetries can be written in terms of electromagnetic responses
\begin{eqnarray}
\sigma_T \left( s, \theta \right) &\equiv& \frac{q_\pi}{k_\gamma} W_{T}\left( s, \theta \right) \label{eq:eq1} \\
\Sigma \left( s, \theta \right) &\equiv& -\frac{W_{S}\left( s, \theta \right)}{W_T\left( s, \theta \right) } \label{eq:eq2} \\
E \left( s, \theta \right) &\equiv& \frac{W_{E}\left( s, \theta \right)}{W_T\left( s, \theta \right)}  \label{eq:eq3} \\
F \left( s, \theta \right) &\equiv& \frac{W_{F}\left( s, \theta \right)}{W_T\left( s, \theta \right)} \: . \label{eq:eq4}
\end{eqnarray}
The responses  $W_T$, $W_{S}$, $W_{E}$, and $W_{F}$
are defined in term of the electromagnetic multipoles up to $D$ waves:
\begin{equation}
\begin{split}
W_{T} =T_0\left( s \right) + T_1\left( s \right) \mathcal{P}_1\left( \theta \right)  
+ T_2\left( s \right) \mathcal{P}_2\left( \theta \right) \: \\
+ T_3\left( s \right) \mathcal{P}_3\left( \theta \right) \: 
+ T_4\left( s \right) \mathcal{P}_4\left( \theta \right)
\end{split} \label{eq:wt}
\end{equation}
\begin{equation}
\begin{split}
W_{S}\left( s, \theta \right) =\left[ S_0 \left( s \right) \right. &+S_1 \left( s \right) \mathcal{P}_1\left(  \theta \right)  \\
&\left. + \: S_2 \left( s \right) \mathcal{P}_2\left( \theta \right)  \: \right] \sin^2 \theta 
\end{split} \label{eq:wtt}
\end{equation}
\begin{equation}
\begin{split}
W_{E} =E_0\left( s \right) + E_1\left( s \right) \mathcal{P}_1\left( \theta \right)  
+ E_2\left( s \right) \mathcal{P}_2\left( \theta \right) \: \\ 
+ E_3\left( s \right) \mathcal{P}_3\left( \theta \right) \:
+ E_4\left( s \right) \mathcal{P}_4\left( \theta \right) 
\end{split} \label{eq:we}
\end{equation}
\begin{equation}
\begin{split}
W_{F}\left( s, \theta \right) =&\left[ F_0 \left( s \right)
+ F_1 \left( s \right) \mathcal{P}_1\left(  \theta \right)  \right. \\
&+ F_2 \left( s \right) \mathcal{P}_2\left( \theta \right) 
\left. + F_3 \left( s \right) \mathcal{P}_3\left( \theta \right) \: \right] \sin \theta  \: ,
\end{split} \label{eq:wf}
\end{equation}
where $P_j \left( \theta \right)$ are the Legendre polynomials in terms of $\cos \theta$ and
\begin{eqnarray}
T_n \left( s \right)&=&\sum_{ij} \text{Re} \{ \: \mathcal{M}^*_i \left( s \right) \: T_n^{ij} \: \mathcal{M}_j \left( s \right) \: \} \label{eq:Tn} \\
S_n \left( s \right)&=&\sum_{ij} \text{Re} \{ \: \mathcal{M}^*_i \left( s \right) \: S_n^{ij} \: \mathcal{M}_j \left( s \right) \: \} \\
E_n \left( s \right)&=&\sum_{ij} \text{Re} \{ \: \mathcal{M}^*_i \left( s \right) \: E_n^{ij} \: \mathcal{M}_j \left( s \right) \: \} \\
F_n \left( s \right)&=&\sum_{ij} \text{Re} \{ \: \mathcal{M}^*_i \left( s \right) \: F_n^{ij} \: \mathcal{M}_j \left( s \right) \: \}
\end{eqnarray}
where
$\mathcal{M}_j \left( s \right) =E_{0+}$, $E_{1+}$, $E_{2+}$, $E_{2-}$, $M_{1+}$, $M_{1-}$, $M_{2+}$, $M_{2-}$.

The coefficients $T_n^{ij}$ are provided in Tables \ref{tab:T0}, \ref{tab:T1}, \ref{tab:T2}, \ref{tab:T3}
 and \ref{tab:T4};
$S_n^{ij}$ in Tables \ref{tab:TT0}, \ref{tab:TT1} and \ref{tab:TT2};
$E_n^{ij}$ in Tables \ref{tab:E0}, \ref{tab:E1}, \ref{tab:E2}, \ref{tab:E3}  and
\ref{tab:E4};
and $F_n^{ij}$ in Tables \ref{tab:F0}, \ref{tab:F1}, \ref{tab:F2} and \ref{tab:F3}. 
The equations necessary to compute the tables can be found in Ref. \cite{DonnellyRaskin}.
For a complete list of polarization observables we refer the reader to Refs. \cite{BDS75,CT97}.

\begin{table}
\caption{$T_0^{ij}$} \label{tab:T0}
\begin{ruledtabular}
\begin{tabular}{|c|c|c|c|c|c|c|c|c|}
 & $E^*_{0+}$ & $E^*_{1+}$ & $E^*_{2+}$  & $E^*_{2-}$  & $M^*_{1+}$ & $M^*_{1-}$ & $M^*_{2+}$  & $M^*_{2-}$ \\\hline
$E_{0+}$ &1&&&&&&&\\
\hline
$E_{1+}$ &&6&&&&&&\\
\hline
$E_{2+}$ &&&18&&&&&\\
\hline
$E_{2-}$ &&&&2&&&&\\
\hline
$M_{1+}$ &&&&&2&&&\\
\hline
$M_{1-}$ &&&&&&1&&\\
\hline
$M_{2+}$ &&&&&&&9&\\
\hline
$M_{2-}$ &&&&&&&&6\\
\end{tabular}
\end{ruledtabular}
\end{table}

\begin{table}
\caption{$T_1^{ij}$} \label{tab:T1}
\begin{ruledtabular}
\begin{tabular}{|c|c|c|c|c|c|c|c|c|}
 & $E^*_{0+}$ & $E^*_{1+}$ & $E^*_{2+}$  & $E^*_{2-}$  & $M^*_{1+}$ & $M^*_{1-}$ & $M^*_{2+}$  & $M^*_{2-}$ \\\hline
$E_{0+}$ &&3&&&1&$-$1&&\\
\hline
$E_{1+}$ &3&&72/5&$-$3/5&&&9/5&$-$9/5\\
\hline
$E_{2+}$ &&72/5&&&&&&\\
\hline
$E_{2-}$ &&$-$3/5&&&1&$-$1&&\\
\hline
$M_{1+}$ &1&&&1&&&27/5&3/5\\
\hline
$M_{1-}$ &$-$1&&&$-$1&&&&3\\
\hline
$M_{2+}$ &&9/5&&&27/5&&&\\
\hline
$M_{2-}$ &&$-$9/5&&&3/5&3&&\\
\end{tabular}
\end{ruledtabular}
\end{table}

\begin{table}
\caption{$T_2^{ij}$} \label{tab:T2}
\begin{ruledtabular}
\begin{tabular}{|c|c|c|c|c|c|c|c|c|}
 & $E^*_{0+}$ & $E^*_{1+}$ & $E^*_{2+}$  & $E^*_{2-}$  & $M^*_{1+}$ & $M^*_{1-}$ & $M^*_{2+}$  & $M^*_{2-}$ \\\hline
$E_{0+}$ &&&6&1&&&3&$-$3\\
\hline
$E_{1+}$ &&3&&&3&$-$3&&\\
\hline
$E_{2+}$ &6&&108/7&$-$12/7&&&36/7&$-$36/7\\
\hline
$E_{2-}$ &1&&$-$12/7&$-$1&&&3&$-$3\\
\hline
$M_{1+}$ &&3&&&$-$1&$-$1&&\\
\hline
$M_{1-}$ &&$-$3&&&$-$1&&&\\
\hline
$M_{2+}$ &3&&36/7&3&&&36/7&9/7\\
\hline
$M_{2-}$ &$-$3&&$-$36/7&$-$3&&&9/7&3\\
\end{tabular}
\end{ruledtabular}
\end{table}

\begin{table}
\caption{$T_3^{ij}$} \label{tab:T3}
\begin{ruledtabular}
\begin{tabular}{|c|c|c|c|c|c|c|c|c|}
 & $E^*_{0+}$ & $E^*_{1+}$ & $E^*_{2+}$  & $E^*_{2-}$  & $M^*_{1+}$ & $M^*_{1-}$ & $M^*_{2+}$  & $M^*_{2-}$ \\\hline
$E_{0+}$ &&&&&&&&\\
\hline
$E_{1+}$ &&&18/5&18/5&&&36/5&$-$36/5\\
\hline
$E_{2+}$ &&18/5&&&6&$-$6&&\\
\hline
$E_{2-}$ &&18/5&&&&&&\\
\hline
$M_{1+}$ &&&6&&&&$-$12/5&$-$18/5\\
\hline
$M_{1-}$ &&&$-$6&&&&$-$3&\\
\hline
$M_{2+}$ &&36/5&&&$-$12/5&$-$3&&\\
\hline
$M_{2-}$ &&$-$36/5&&&$-$18/5&&&\\
\end{tabular}
\end{ruledtabular}
\end{table}

\begin{table}
\caption{$T_4^{ij}$} \label{tab:T4}
\begin{ruledtabular}
\begin{tabular}{|c|c|c|c|c|c|c|c|c|}
 & $E^*_{0+}$ & $E^*_{1+}$ & $E^*_{2+}$  & $E^*_{2-}$  & $M^*_{1+}$ & $M^*_{1-}$ & $M^*_{2+}$  & $M^*_{2-}$ \\\hline
$E_{0+}$ &&&&&&&&\\
\hline
$E_{1+}$ &&&&&&&&\\
\hline
$E_{2+}$ &&&18/7&54/7&&&90/7&$-$90/7\\
\hline
$E_{2-}$ &&&54/7&&&&&\\
\hline
$M_{1+}$ &&&&&&&&\\
\hline
$M_{1-}$ &&&&&&&&\\
\hline
$M_{2+}$ &&&90/7&&&&$-$36/7&$-$72/7\\
\hline
$M_{2-}$ &&&$-$90/7&&&&$-$72/7&\\
\end{tabular}
\end{ruledtabular}
\end{table}

\begin{table}
\caption{$S_0^{ij}$} \label{tab:TT0}
\begin{ruledtabular}
\begin{tabular}{|c|c|c|c|c|c|c|c|c|}
 & $E^*_{0+}$ & $E^*_{1+}$ & $E^*_{2+}$  & $E^*_{2-}$  & $M^*_{1+}$ & $M^*_{1-}$ & $M^*_{2+}$  & $M^*_{2-}$ \\\hline
$E_{0+}$ & &&3/2&3/2&&&$-$3/2&3/2\\
\hline
$E_{1+}$ & &$-$9/2 &&&3/2&$-$3/2&&\\
\hline
$E_{2+}$ &3/2 & &$-$24&6&&&6&$-$6\\
\hline
$E_{2-}$ &3/2  & &6&$-$3/2&&&$-$3/2&3/2\\
\hline
$M_{1+}$ & &3/2 &&&3/2&3/2&&\\
\hline
$M_{1-}$ & &$-$3/2 &&&3/2&&&\\
\hline
$M_{2+}$ &$-$3/2 & &6&3/2&&&12&21/2\\
\hline
$M_{2-}$  &3/2 & &$-$6&3/2&&&21/2&9/2\\
\end{tabular}
\end{ruledtabular}
\end{table}

\begin{table}
\caption{$S_1^{ij}$}  \label{tab:TT1} 
\begin{ruledtabular}
\begin{tabular}{|c|c|c|c|c|c|c|c|c|}
 & $E^*_{0+}$ & $E^*_{1+}$ & $E^*_{2+}$  & $E^*_{2-}$  & $M^*_{1+}$ & $M^*_{1-}$ & $M^*_{2+}$  & $M^*_{2-}$ \\\hline
$E_{0+}$ & &&&&&&&\\
\hline
$E_{1+}$ & & &$-$27/2&9&&&&\\
\hline
$E_{2+}$ & &$-$27/2 &&&15/2&$-$15/2&&\\
\hline
$E_{2-}$ &  &9 &&&&&&\\
\hline
$M_{1+}$ & & &15/2&&&&6&9\\
\hline
$M_{1-}$ & & &$-$15/2&&&&15/2&\\
\hline
$M_{2+}$ & & &&&6&15/2&&\\
\hline
$M_{2-}$  & & &&&9&&&\\
\end{tabular}
\end{ruledtabular}
\end{table}

\begin{table}
\caption{$S_2^{ij}$} \label{tab:TT2}
\begin{ruledtabular}
\begin{tabular}{|c|c|c|c|c|c|c|c|c|}
 & $E^*_{0+}$ & $E^*_{1+}$ & $E^*_{2+}$  & $E^*_{2-}$  & $M^*_{1+}$ & $M^*_{1-}$ & $M^*_{2+}$  & $M^*_{2-}$ \\\hline
$E_{0+}$ & &&&&&&&\\
\hline
$E_{1+}$ & & &&&&&&\\
\hline
$E_{2+}$ & & &$-$30&45/2&&&15/2&$-$15/2\\
\hline
$E_{2-}$ &  & &45/2&&&&&\\
\hline
$M_{1+}$ & & &&&&&&\\
\hline
$M_{1-}$ & & &&&&&&\\
\hline
$M_{2+}$ & & &15/2&&&&15&30\\
\hline
$M_{2-}$  & & &$-$15/2&&&&30&\\
\end{tabular}
\end{ruledtabular}
\end{table}

\begin{table}
\caption{$E_0^{ij}$} \label{tab:E0}
\begin{ruledtabular}
\begin{tabular}{|c|c|c|c|c|c|c|c|c|}
& $E_{0+}^*$ & $E_{1+}^*$ & $E_{2+}^*$  & $E_{2-}^*$  & $M_{1+}^*$ & $M_{1-}^*$ & $M_{2+}^*$  & $M_{2-}^*$ \\ 
 \hline
$E_{0+}$ &1&&&&&&&\\
\hline
$E_{1+}$ &&3&&&3&&&\\
\hline
$E_{2+}$ &&&6&&&&12&\\
\hline
$E_{2-}$ &&&&$-$1&&&&$-$3\\
\hline
$M_{1+}$ &&3&&&$-$1&&&\\
\hline
$M_{1-}$ &&&&&&1&&\\
\hline
$M_{2+}$ &&&12&&&&$-$3&\\
\hline
$M_{2-}$ &&&&$-$3&&&&3\\
\end{tabular}
\end{ruledtabular}
\end{table}

\begin{table}
\caption{$E_1^{ij}$} \label{tab:E1}
\begin{ruledtabular}
\begin{tabular}{|c|c|c|c|c|c|c|c|c|}
& $E_{0+}^*$ & $E_{1+}^*$ & $E_{2+}^*$  & $E_{2-}^*$  & $M_{1+}^*$ & $M_{1-}^*$ & $M_{2+}^*$  & $M_{2-}^*$ \\ 
 \hline
$E_{0+}$ &&3&&&1&$-$1&&\\
\hline
$E_{1+}$ &3&&36/5&6/5&&&9&\\
\hline
$E_{2+}$ &&36/5&&&36/5&&&\\
\hline
$E_{2-}$ &&6/5&&&$-$4/5&$-$1&&\\
\hline
$M_{1+}$ &1&&36/5&$-$4/5&&&$-$9/5&$-$6/5\\
\hline
$M_{1-}$ &$-$1&&&$-$1&&&&3\\
\hline
$M_{2+}$ &&9&&&$-$9/5&&&\\
\hline
$M_{2-}$ &&&&&$-$6/5&3&&\\

\end{tabular}
\end{ruledtabular}
\end{table}

\begin{table}
\caption{$E_2^{ij}$} \label{tab:E2}
\begin{ruledtabular}
\begin{tabular}{|c|c|c|c|c|c|c|c|c|}
& $E_{0+}^*$ & $E_{1+}^*$ & $E_{2+}^*$  & $E_{2-}^*$  & $M_{1+}^*$ & $M_{1-}^*$ & $M_{2+}^*$  & $M_{2-}^*$ \\ 
 \hline
$E_{0+}$ &&&6&1&&&3&$-$3\\
\hline
$E_{1+}$ &&6&&&&$-$3&&\\
\hline
$E_{2+}$ &6&&12&24/7&&&60/7&\\
\hline
$E_{2-}$ &1&&24/7&2&&$-$15/7&&\\
\hline
$M_{1+}$ &&&&&2&$-$1&&\\
\hline
$M_{1-}$ &&$-$3&&$-$1&&&&\\
\hline
$M_{2+}$ &3&&60/7&$-$15/7&&&12/7&$-$27/7\\
\hline
$M_{2-}$ &$-$3&&&&&&$-$27/7&6\\
\end{tabular}
\end{ruledtabular}
\end{table}

\begin{table}
\caption{$E_3^{ij}$} \label{tab:E3}
\begin{ruledtabular}
\begin{tabular}{|c|c|c|c|c|c|c|c|c|}
& $E_{0+}^*$ & $E_{1+}^*$ & $E_{2+}^*$  & $E_{2-}^*$  & $M_{1+}^*$ & $M_{1-}^*$ & $M_{2+}^*$  & $M_{2-}^*$ \\ 
 \hline
$E_{0+}$ &&&&&&&&\\
\hline
$E_{1+}$ &&&54/5&9/5&&&&$-$9\\
\hline
$E_{2+}$ &&54/5&&&$-$6/5&$-$6&&\\
\hline
$E_{2-}$ &&9/5&&&9/5&&&\\
\hline
$M_{1+}$ &&&$-$6/5&9/5&&&24/5&$-$9/5\\
\hline
$M_{1-}$ &&&$-$6&&&&$-$3&\\
\hline
$M_{2+}$ &&&&&24/5&$-$3&&\\
\hline
$M_{2-}$ &&$-$9&&&$-$9/5&&&\\
\end{tabular}
\end{ruledtabular}
\end{table}

\begin{table}
\caption{$E_4^{ij}$} \label{tab:E4}
\begin{ruledtabular}
\begin{tabular}{|c|c|c|c|c|c|c|c|c|}
& $E_{0+}^*$ & $E_{1+}^*$ & $E_{2+}^*$  & $E_{2-}^*$  & $M_{1+}^*$ & $M_{1-}^*$ & $M_{2+}^*$  & $M_{2-}^*$ \\ 
 \hline
$E_{0+}$ &&&&&&&&\\
\hline
$E_{1+}$ &&&&&&&&\\
\hline
$E_{2+}$ &&&18&18/7&&&$-$18/7&$-$18\\
\hline
$E_{2-}$ &&&18/7&&&&36/7&\\
\hline
$M_{1+}$ &&&&&&&&\\
\hline
$M_{1-}$ &&&&&&&&\\
\hline
$M_{2+}$ &&&$-$18/7&36/7&&&72/7&$-$36/7\\
\hline
$M_{2-}$ &&&$-$18&&&&$-$36/7&\\
\end{tabular}
\end{ruledtabular}
\end{table}

\begin{table}
\caption{$F_0^{ij}$} \label{tab:F0}
\begin{ruledtabular}
\begin{tabular}{|c|c|c|c|c|c|c|c|c|}
& $E_{0+}^*$ & $E_{1+}^*$ & $E_{2+}^*$  & $E_{2-}^*$  & $M_{1+}^*$ & $M_{1-}^*$ & $M_{2+}^*$  & $M_{2-}^*$ \\ 
 \hline
$E_{0+}$ &&$-$3/2&&&3/2&&&\\
\hline
$E_{1+}$ &$-$3/2&&$-$15/2&&&&15/2&\\
\hline
$E_{2+}$ &&$-$15/2&&&$-$5/2&1&&\\
\hline
$E_{2-}$ &&&&&&$-$3/2&&\\
\hline
$M_{1+}$ &3/2&&$-$5/2&&&&5/2&\\
\hline
$M_{1-}$ &&&1&$-$3/2&&&$-$1&$-$3/2\\
\hline
$M_{2+}$ &&15/2&&&5/2&$-$1&&\\
\hline
$M_{2-}$ &&&&&&$-$3/2&&\\
\end{tabular}
\end{ruledtabular}
\end{table}

\begin{table}
\caption{$F_1^{ij}$} \label{tab:F1}
\begin{ruledtabular}
\begin{tabular}{|c|c|c|c|c|c|c|c|c|}
& $E_{0+}^*$ & $E_{1+}^*$ & $E_{2+}^*$  & $E_{2-}^*$  & $M_{1+}^*$ & $M_{1-}^*$ & $M_{2+}^*$  & $M_{2-}^*$ \\ 
 \hline
$E_{0+}$ &&&$-$6&3/2&&&6&3/2\\
\hline
$E_{1+}$ &&$-$9&&&3&3/2&&\\
\hline
$E_{2+}$ &$-$6&&$-$36&$-$3/2&&&9&9/2\\
\hline
$E_{2-}$ &3/2&&$-$3/2&3&&&3/2&$-$3\\
\hline
$M_{1+}$ &&3&&&3&$-$3/2&&\\
\hline
$M_{1-}$ &&3/2&&&$-$3/2&&&\\
\hline
$M_{2+}$ &6&&9&3/2&&&18&$-$9/2\\
\hline
$M_{2-}$ &3/2&&9/2&$-$3&&&$-$9/2&$-$9\\
\end{tabular}
\end{ruledtabular}
\end{table}

\begin{table}
\caption{$F_2^{ij}$} \label{tab:F2}
\begin{ruledtabular}
\begin{tabular}{|c|c|c|c|c|c|c|c|c|}
& $E_{0+}^*$ & $E_{1+}^*$ & $E_{2+}^*$  & $E_{2-}^*$  & $M_{1+}^*$ & $M_{1-}^*$ & $M_{2+}^*$  & $M_{2-}^*$ \\ 
 \hline
$E_{0+}$ &&&&&&&&\\
\hline
$E_{1+}$ &&&$-$39/2&3&&&6&9\\
\hline
$E_{2+}$ &&$-$39/2&&&11/2&5&&\\
\hline
$E_{2-}$ &&3&&&3&&&\\
\hline
$M_{1+}$ &&&11/2&3&&&8&$-$3\\
\hline
$M_{1-}$ &&&5&&&&$-$5&\\
\hline
$M_{2+}$ &&6&&&8&$-$5&&\\
\hline
$M_{2-}$ &&9&&&$-$3&&&\\
\end{tabular}
\end{ruledtabular}
\end{table}

\begin{table}
\caption{$F_3^{ij}$} \label{tab:F3}
\begin{ruledtabular}
\begin{tabular}{|c|c|c|c|c|c|c|c|c|}
& $E_{0+}^*$ & $E_{1+}^*$ & $E_{2+}^*$  & $E_{2-}^*$  & $M_{1+}^*$ & $M_{1-}^*$ & $M_{2+}^*$  & $M_{2-}^*$ \\ 
 \hline
$E_{0+}$ &&&&&&&&\\
\hline
$E_{1+}$ &&&&&&&&\\
\hline
$E_{2+}$ &&&$-$36&9/2&&&9&45/2\\
\hline
$E_{2-}$ &&&9/2&&&&9&\\
\hline
$M_{1+}$ &&&&&&&&\\
\hline
$M_{1-}$ &&&&&&&&\\
\hline
$M_{2+}$ &&&9&9&&&18&$-$9\\
\hline
$M_{2-}$ &&&45/2&&&&$-$9&\\
\end{tabular}
\end{ruledtabular}
\end{table}


\begin{thebibliography}{99}
\bibitem{book} J. F. Donoghue, E. Golowich, and B. R. Holstein,
\textit{Cambridge Monographs in Particle Physics, Nuclear Physics and Cosmology Vol. 2:
Dynamics of the Standard Model}
 (Cambridge University Press, Cambridge, 1992).
\bibitem{CHPT91}  V. Bernard, N. Kaiser, J. Gasser, and U.-G. Mei{\ss}ner,
Phys. Lett.  \textbf{B268}, 291 (1991);
V. Bernard, N. Kaiser, and U.-G. Mei{\ss}ner, Nucl. Phys. \textbf{B383}, 442 (1992).
\bibitem{CHPT96} V. Bernard, N. Kaiser, and U.-G. Mei{\ss}ner, Z. Phys. \textbf{C70}, 483 (1996).
\bibitem{CHPT01} V. Bernard, N. Kaiser, and U.-G. Mei{\ss}ner, Eur. Phys. J. \textbf{A11}, 209 (2001).
\bibitem{AB-Delta} A. M. Bernstein and S. Stave, Few Body Syst. \textbf{41}, 83 (2007).
\bibitem{AB-fits} A. M. Bernstein,  E. Shuster, R. Beck, M. Fuchs, B. Krusche, H. Merkel, and H. Str\"oher,
Phys. Rev. C \textbf{55}, 1509 (1997).
\bibitem{dwavesindelta} R. Beck \textit{et al.}, Phys. Rev. C \textbf{61}, 035204 (2000). 
\bibitem{SAID} R.A. Arndt, W.J. Briscoe, I.I. Strakovsky, and R.L. Workman,
Phys. Rev. C 66, 055213 (2002),
R.A. Arndt, I.I. Strakovsky, and R.L. Workman, Int. J. Mod. Phys. A 18, 449 (2003),
SAID database, \url{http://gwdac.phys.gwu.edu}.
\bibitem{Schmidt} A. Schmidt \textit{et al.}, Phys. Rev. Lett. \textbf{87}, 232501 (2001).
\bibitem{AB-review} A.M. Bernstein, M.W. Ahmed, S. Stave, Y.K. Wu, H.R. Weller, Annu. Rev. Nucl. Part. Sci. 59, 115 (2009).
\bibitem{Peccei} R.D. Peccei, Phys. Rev. \textbf{181}, 1902 (1969).
\bibitem{DMT} S.S. Kamalov, G.-Y. Chen, S.N. Yang, D. Drechsel, and L. Tiator,
Phys. Lett. \textbf{B522}, 27 (2001).
\bibitem{FBD09} C. Fern\'andez-Ram\'{\i}rez, A. M. Bernstein, and T. W. Donnelly,
Phys. Lett. \textbf{B679}, 41 (2009).
\bibitem{DonnellyRaskin} T. W. Donnelly and A. S. Raskin, Ann. Phys. (N.Y.) \textbf{169}, 247 (1986);
A. S. Raskin and T. W. Donnelly, Ann. Phys. (N.Y.) \textbf{191}, 78 (1989).
\bibitem{BDS75} I. S. Barker, A. Donnachie, and J. K. Storrow, Nucl. Phys. \textbf{B95}, 347 (1975).
\bibitem{CT97} W. T. Chiang and F. Tabakin, Phys. Rev. C \textbf{55}, 2054 (1997).
\bibitem{Knochlein95} G. Knochlein, D. Drechsel, and L. Tiator,  Z. Phys. \textbf{A352}, 327 (1995). 
\bibitem{FMU06} C. Fern\'andez-Ram\'{\i}rez, E. Moya de Guerra, and J. M. Ud\'{\i}as, 
Ann. Phys. (N.Y.) \textbf{321}, 1408 (2006); Phys. Lett. \textbf{B660}, 188 (2008).
\bibitem{MMD} P. Mergell, U.-G. Mei{\ss}ner, and D. Drechsel, Nucl. Phys. \textbf{A596}, 367 (1996).
\bibitem{Belushkin} M.A. Belushkin, H.-W. Hammer, and U.-G. Mei{\ss}ner,
Phys. Rev. C \textbf{75} 035202 (2007).
\bibitem{NAG} Numerical Algorithms Group Ltd.,
 Wilkinson House, Jordan Hill Road, Oxford OX2-8DR, UK,
\url{http://www.nag.co.uk}
\bibitem{OPTIMIZACION} C. Winkler and H. M. Hofmann,
Phys. Rev. C \textbf{55}, 684 (1997);
I. Golovkin \textit{et al.}, Phys. Rev. Lett. \textbf{88}, 045002 (2002);
S. Janssen, D. G. Ireland, and J. Ryckebusch,
Phys. Lett.  \textbf{B562}, 51 (2003);
D. G. Ireland, S. Janssen, and J. Ryckebusch, 
Nucl. Phys.  \textbf{A740}, 147 (2004);
B. C. Allanach, D. Grellscheid, and F. Quevedo,
J. High Energy Phys. JHEP \textbf{07}, 069 (2004).
\bibitem{PRC08} C. Fern\'andez-Ram\'{\i}rez, E. Moya de Guerra, A. Ud\'{\i}as, and J. M. Ud\'{\i}as,
Phys. Rev. C \textbf{77}, 065212 (2008).
\bibitem{Goldberg} D. E. Goldberg, \textit{Genetic Algorithms in Search, 
Optimization \& Machine Learning} (Addison Wesley, Reading MA, 1989);
L. Davis, \textit{Handbook of Genetic Algorithms}
(Van Nostrand Reinhold, New York NY, 1991);
Z. Michalewicz, \textit{Genetic Algorithms+Data 
Structures=Evolution Programs} (Springer, Berlin-Heidelberg-New York, 1999);
K. Deb, \textit{Multi-Objective Optimization Using 
Evolutionary Algorithms} (Wiley, New York NY, 2002).
\bibitem{AB-lq} A. M. Bernstein, Phys. Lett. \textbf{B442}, 20 (1998).
\bibitem{Anant} B. Ananthanarayan, Phys. Lett. \textbf{B634}, 391 (2006).
\bibitem{FW} K. M. Watson, Phys. Rev. \textbf{95}, 228 (1954);
E. Fermi, Suppl. Nuovo Cimento \textbf{2}, 17 (1955).
\bibitem{VKM2005} V. Bernard, B. Kubis, and U.-G. Mei{\ss}ner, Eur. Phys. J. \textbf{A25}, 419 (2005);
U.-G. Mei{\ss}ner, private communication.
\bibitem{Hornidge} D. Hornidge (spokeperson) \textit{et al.}, 
Mainz Exp. A2/6-03, Measurement of the Photon Asymmetry in Neutral Pion
Production from the Proton near Threshold (2008).
\bibitem{MeissnerNPA00} P. B\"uttiker and U.-G. Mei{\ss}ner, Nucl. Phys. \textbf{A668}, 97 (2000).
\bibitem{Gotta} D. Gotta,  AIP Conf. Proc. \textbf{1037}, 162 (2008).
\bibitem{s-pred} V. Bernard, N. Kaiser, and U.-G. Mei{\ss}ner,  Phys. Lett. \textbf{B309}, 421 (1993).
\bibitem{Korkmaz} E. Korkmaz \textit{et al.}, Phys. Rev. Lett. \textbf{83}, 3609 (1999).
\bibitem{KR-chiral} V. Bernard, N. Kaiser, and U.-G. Mei{\ss}ner, Phys. Lett. \textbf{B383}, 116 (1996).
\bibitem{Pasquini} B. Pasquini, D. Drechsel, and L. Tiator, Eur. Phys. J. \textbf{A23}, 279 (2005).
\bibitem{Fubini} S. Fubini, G. Furlan, and C. Rossetti, Nuovo Cimento \textbf{40}, 1171 (1965).
\bibitem{MAID} D. Drechsel, O. Hanstein, S.S. Kamalov, and L. Tiator, Nucl. Phys. A 645, 145 (1999);
\url{http://www.kph.uni-mainz.de/MAID/}.
\bibitem{MAMI-Bernstein} 
A. M. Bernstein, W. Deconinck, D. Hornidge, M. Ostrick (co-spokesmen) \textit{et al.}, Mainz Exp. A2/10-2009, Measurement of Polarized Target and Beam Asymmetries in Pion Photo-Production on the Proton: Test of Chiral Dynamics (2009).
\end{thebibliography}
\end{document}